\documentclass[floatfix,aps,prb,twocolumn,showpacs,nofootinbib,superscriptaddress]{revtex4-2}
\usepackage{graphicx}
\usepackage{epsfig}

\usepackage{pdfpages}

\makeatletter
\AtBeginDocument{\let\LS@rot\@undefined}
\makeatother

\usepackage{amsmath}
\usepackage{amsfonts}
\usepackage{amssymb}
\usepackage{bm}
\usepackage{mathtools}

\usepackage{natbib}
\renewcommand\harvardurl[1]{\textbf{URL:} \url{#1}}
\usepackage{hyperref}
\usepackage{balance}
\usepackage{dsfont}
\usepackage{lipsum}
\usepackage{epstopdf}
\usepackage{comment}
\usepackage{chngcntr}

\usepackage{subfig}

\usepackage{braket}
\usepackage{tikz}
\usepackage{circuitikz}
\usetikzlibrary{arrows, shapes}
\usepackage{pgfplots}
\pgfplotsset{compat=1.14}
\usepackage [english]{babel}
\usepackage [autostyle, english = american]{csquotes}
\MakeOuterQuote{"}

\usepackage{chngcntr}

\begin{document}

\author{Ashwin Gopal}
\affiliation{Complex Systems and Statistical Mechanics, Department of Physics and Materials Science,
University of Luxembourg, L-1511 Luxembourg, Luxembourg}

\author{Massimiliano Esposito}
\affiliation{Complex Systems and Statistical Mechanics, Department of Physics and Materials Science,
University of Luxembourg, L-1511 Luxembourg, Luxembourg}

\author{Nahuel Freitas}
\affiliation{Universidad de Buenos Aires, Facultad de Ciencias Exactas y Naturales, Departamento de Física. Buenos Aires, Argentina}

\title{Thermodynamic cost of precise timekeeping in an electronic underdamped clock}
\date{\today}

\begin{abstract}
Clocks are inherently out-of-equilibrium because, due to friction, they constantly consume free energy to keep track of time. The Thermodynamic Uncertainty Relation (TUR) quantifies the trade-off between the precision of any time-antisymmetric observable and entropy production. In the context of clocks, the TUR implies that a minimum entropy production is needed in order to achieve a certain level of precision in timekeeping. But the TUR has only been proven for overdamped systems. Recently, a toy model of a classical underdamped pendulum clock was proposed that violated this relation (Phys. Rev. Lett. 128, 130606), thus demonstrating that the TUR does not hold for underdamped dynamics. We propose an electronic implementation of such a clock, using a resistor-inductor-capacitor (RLC) circuit and a biased CMOS inverter (NOT gate), which can work at different scales. We find that in the nanoscopic single-electron regime of the circuit, we essentially recover the toy model violating the TUR bound. However, in different macroscopic regimes of the circuit, we show that the TUR bound is restored and analyze the thermodynamic efficiency of timekeeping.
\end{abstract}

\maketitle

\section{Introduction}
Timekeeping has a long history, dating back to ancient Mesopotamia around 2000 B.C., where the earliest forms of clocks emerged through the use of sundials and astronomical clocks \cite{brouwer1951accurate, jespersen1999sundials}. The need for more practical time-keeping at finer intervals, without the need for celestial tables, later led to the invention of portable clocks, like sand glasses, water clocks, candle clocks, etc. All of these clocks utilize the irreversible relaxation dynamics of a system to provide a time reference. However, the limited run-time and accuracy motivated better designs for clocks. The greatest leap in timekeeping came from the theoretical understanding of the oscillatory dynamics in a pendulum, where the time period for small oscillations is independent of its amplitude. In 1656, Christian Huygens designed the first pendulum clock using a mechanical linkage to move the clock's loaded hand forward based on the periodic motion of a pendulum \cite{thomson1842time}. This mechanical linkage, called the escapement, is still a widely used mechanism for precise timekeeping in modern mechanical watches. The quest for increased precision of timekeeping has led us to the current standard of time using atomic clocks. These clocks use the resonant frequency of atoms, and can reach uncertainties of 1 second in 300 million years (i.e. relative uncertainty of $10^{-16}$) \cite{li2011improved}. \par 


Since all realistic clocks undergo friction, precise timekeeping requires an input power source such as a battery, physical winding, chemical potential difference, electromagnetic driving, etc. Clocks, like thermal machines, are thus inherently non-equilibrium systems. The fundamental connection between thermodynamics and the efficiency of clocks has recently been explored using the tools of stochastic thermodynamics \cite{cao2015free, marsland2019thermodynamic, barato2016cost,erker2017autonomous,milburn2020thermodynamics,pearson2021measuring,pietzonka2022classical}. Autonomous clocks, both classical \cite{pearson2021measuring, barato2016cost} and quantum \cite{erker2017autonomous}, have been used to derive a linear relationship between the maximum precision in timekeeping and the dissipation in the system. These results are consistent with the trade-off relations observed in some non-equilibrium systems, known as the thermodynamic uncertainty relations (TURs) \cite{barato2016cost, marsland2019thermodynamic}.

As originally formulated, the TUR establishes an upper bound on the precision of an observable based on the entropy production of the system \cite{barato2015thermodynamic, horowitz2020thermodynamic, falasco2020unifying}. For long times, the product of the entropy production and the constant temperature $T$ of the environment corresponds to the dissipated heat. Hence, in the context of clocks, the TUR implies that a minimum dissipation is needed for a given precision in timekeeping. These bounds have been proven for Markov jump and diffusion processes, which are overdamped systems where the momenta degrees of freedom have relaxed \cite{horowitz2017proof, dieball2023direct}. Most of the clock dynamics previously explored, in this context, are overdamped \cite{barato2015thermodynamic,marsland2019thermodynamic,erker2017autonomous,helms2022stochastic} and hence display a minimum thermodynamic cost.

However, the original TUR has not yet been proven for underdamped systems in which the inertial dynamics is retained. Several modified versions of TUR, including kinetic quantities, such as dynamical activity, have been proposed for underdamped systems \cite{van2019uncertainty, lee2021universal}. Similarly, finite time versions, involving only the entropy production, have been conjectured which converge to the original TUR in the long-time limit \cite{fischer2020free}. Underdamped systems in equilibrium can have oscillatory relaxation dynamics that, if driven, can provide sustained oscillations. The inherent thermal noise in these systems can serve as the driving force to produce stochastic oscillations, which incur no additional thermodynamic cost. Hence, the thermodynamic cost of timekeeping using these stochastic underdamped oscillations should ideally provide an energy-efficient alternative compared to the overdamped counterparts. 

Recently, a toy model clock exploiting the escapement mechanism has been proposed, that violates the original TUR bound \cite{pietzonka2022classical}. This clock comprises two key components: a stochastic oscillator, which is modeled as an underdamped harmonic oscillator, and a discrete counter, modeled as a Markov jump process. The counter has a thermodynamic bias, that makes it to move preferably in a forward direction and mimics the motion of the `hand' in a classical clock. The novelty of this toy model is to couple the counter-jumps with the dynamics of the stochastic oscillator. Specifically, the counter advances only during each zero crossing of the oscillator's position, which occurs periodically with a precision dependent only on the damping and the temperature of the environment. This coupling mechanism between an underdamped dynamics and a discrete counter effectively emulates the concept of an escapement, known for improving the precision of mechanical clocks. Since equilibrium thermal noise is used to probe the natural frequency of the underdamped oscillator, it provides a natural timescale without any additional thermodynamic cost, thus enabling the violation of the TUR bound. Fundamentally, this also implies that precise clocks can be ideally built with higher thermodynamic efficiency than the one established by the TUR.

Motivated by those intriguing findings, we extend the idea to the realm of electronic circuits, aiming to explore the potential realization of TUR violations in a practical setting.  We propose a simple electronic implementation of the escapement clock, using an equilibrium RLC oscillator and a biased CMOS inverter. A crucial feature of our model is that it works at different scales, which is related to the size of the electronic components. In the deep nanoscopic regime for both the oscillator and the counter, the counter dynamics only involves states with a few electrons and the transition rates can be considered to depend only on the sign of the oscillator voltage. Thus, in that single-electron regime of operation, our model reduces to the toy model mentioned above, where the TUR is indeed violated. However, for all the other operation regimes the TUR bound is restored, and we highlight the key ingredients needed for it to be valid. 

\par

The paper is structured as follows, in Sec.~\ref{sec: Model}, we describe the working principle and the stochastic modelling of our underdamped clock circuit. In Sec.~\ref{sec: Analysis}, we define an uncertainty product to quantify the thermodynamic efficiency of the clock. We will also describe the dynamical consequences of scaling down the size of its components to achieve the single electron regime of the circuit. In Sec.~\ref{sec: Results}, we will present the numerical and/or analytical results of the uncertainty product of the clock at different physical scales of its components, and compare it with the TUR bound. In Sec.~\ref{sec: coarse_grained_derivation}, we provide the derivation for a general expression of the uncertainty product in regimes when the voltage at the RLC can be coarse-grained.

\section{Model}
\label{sec: Model}

\begin{figure*}[ht!]
    \centering
    \includegraphics[scale=0.05]{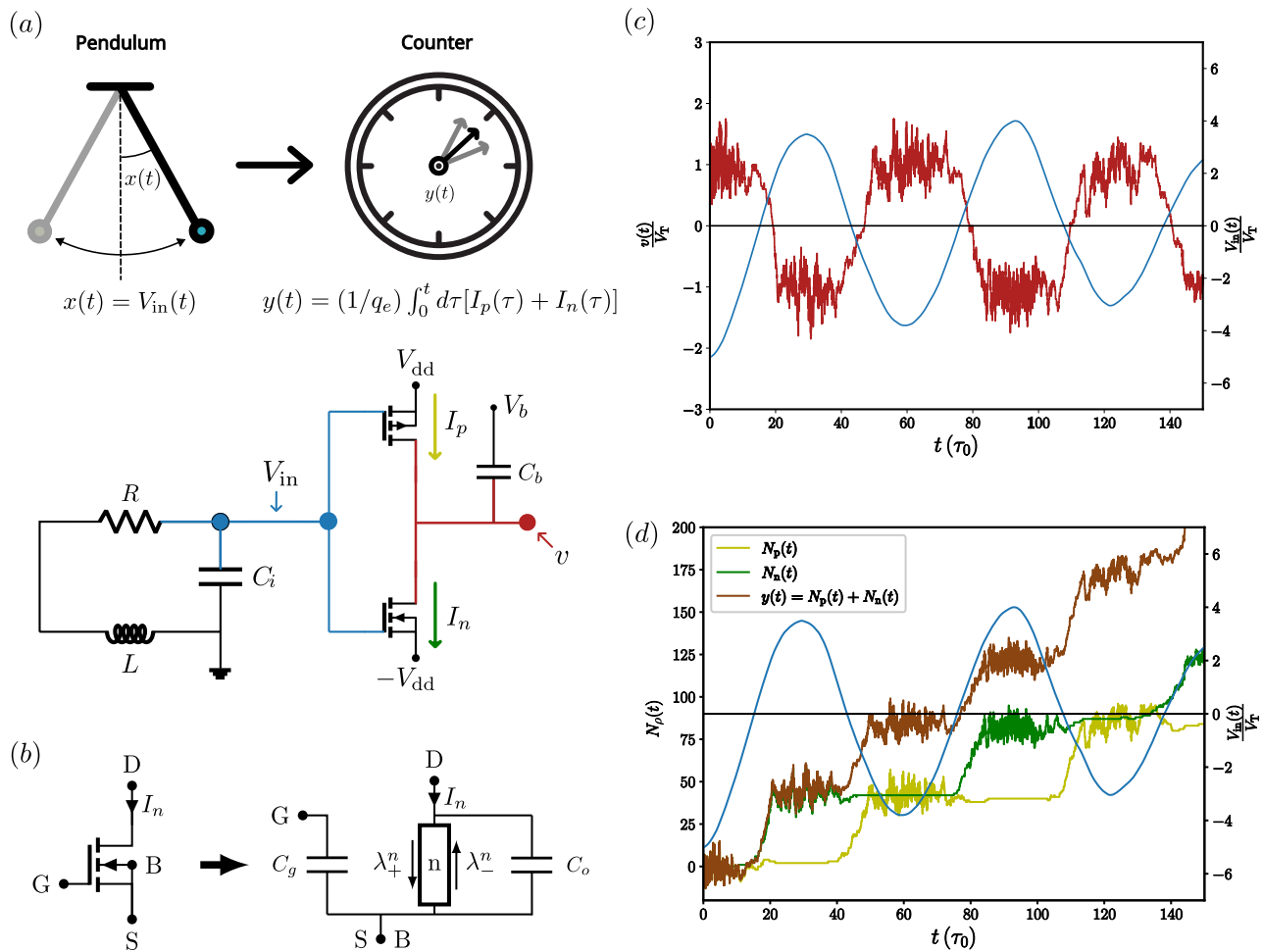}
    \caption{(a) Circuit diagram of the escapement clock. It consists of two parts: an RLC circuit playing the role of a pendulum producing oscillations and described by the voltage $V_{\text{in}}(t)$, and a biased CMOS inverter (NOT gate), with instantaneous electric currents $I_{p/n}(\tau)$, playing the role of a discrete counter of the passage of time and described by the counter observable $y(t)$. (b) Simplified modeling of an nMOS transistor as a conduction channel between its drain (D) and source (S) terminals, with associated Poisson rates $\lambda_n^{\pm}$. The gate-body (G-B) interface is represented as a capacitor $C_g$, and another capacitor $C_o$ takes into account the output capacitance. (c) A sample stochastic trajectory of the input voltage $V_{\text{in}}(t)$ (blue) and the output voltage $v(t)$ (red) in the escapement clock circuit (time in units of $\tau_0=(q_e/I_0)\exp[V_{\rm{th}}/(V_T)]$). (d) The corresponding stochastic dynamics of the integrated currents, $N_{p/n}(t) =(1/q_e)\int_0^t  I_{p/n}(\tau)d\tau$, flowing through p/nMOS (yellow/green), respectively, and the counter-observable $y(t)= N_{p}(t) + N_{n}(t)$  (brown) for the same input voltage trajectory $V_{\text{in}}(t)$ (blue). Parameters: $\omega \equiv 1/\sqrt{LC_{\rm{in}}} = 0.1 \tau_0^{-1}, \gamma/\omega \equiv R\sqrt{C_{\rm{in}}/L} = 0.1, v_e^{\rm{in}}/V_T = 2, v_e^{\rm{out}}/V_T = 0.1 ,   V_{\rm{dd}}/V_T=1$ and $V_{\rm{b}}/V_T = 2$.}
    \label{fig:Circuit}
\end{figure*}
Our circuit implementation of the classical escapement clock consists of an RLC circuit coupled to a biased CMOS inverter (see Fig.~\ref{fig:Circuit}-(a)).  The CMOS inverter is the modern implementation of the logical NOT gate \cite{wang2006sub}. It is made up of a pMOS (top) and an nMOS (bottom) transistor with common gate and drain terminals (at voltages $V_\text{in}$ and $v$ in Fig. \ref{fig:Circuit}-(a)), which act as input and output terminals in logical operations, respectively. The inverter is powered by a fixed potential difference between the source terminals of both transistors, $\Delta V = 2V_\text{dd}$. In our circuit, the output voltage is also biased by a fixed voltage $V_b$ through a capacitive coupling. Therefore, the voltage at the input $V_\text{in}$ and the output terminal $v$ are the two degrees of freedom of our circuit. We consider the operation of this circuit at a finite temperature $T$ with an associated thermal voltage $V_{T} \equiv k_b T/q_e$, where $q_e$ is the positive electron charge.

In this clock circuit, the RLC circuit plays the role of a pendulum, as it is the dynamical equivalent of an underdamped oscillator. The RLC circuit is at thermal equilibrium since it is not powered by any voltage source and since there is also no current flow through the input terminal of the inverter due to the insulating nature of the gate terminal. Nevertheless, the white thermal noise due to the resistor will excite it and produce stochastic oscillations with a coherence time that will depend on the temperature and damping. These stochastic voltage oscillations in the RLC will be the input signal for the inverter, which will modulate the conductivity of the two MOS transistors. This makes the integrated current to have increments at a pace regulated by the natural frequency of the RLC. 
Thus, the biased CMOS inverter acts as the counter, and the integrated current through it is the observable that tracks the passage of time.


In Fig.~\ref{fig:Circuit}. (c), we plot a typical stochastic trajectory for the input voltage $V_{\rm{in}}(t)$ and the output voltage $v(t)$. When $V_{\text{in}}(t)<0$, the conduction through the pMOS transistor is enhanced, while it is reduced for the nMOS transistor, and therefore the output voltage $v$ quickly relaxes to $V_{\rm{dd}}$. Similarly, when the input voltage $V_{\text{in}}(t)>0$, the situation is reversed and the output voltage $v$ relaxes quickly to $-V_{\rm{dd}}$. As seen in Fig.~\ref{fig:Circuit}. (d), for every positive zero crossing ($V_{\text{in}}<0 \to V_{\text{in}}>0$), there will be a net number of charges flowing through the nMOS transistor ($N_{n}(t)=(1/q_e)\int_0^t I_{n}(\tau) d\tau $) from the output terminal to its source terminal to account for the change in output voltage ($+V_{\rm{dd}} \to -V_{\rm{dd}} $). Similarly, for every negative zero crossing ($V_{\text{in}}>0 \to V_{\text{in}}<0$), there will be a net number of charges flowing through the pMOS transistor ($N_{p}(t)=(1/q_e)\int_0^t I_{p}(\tau) d\tau $) from its source terminal to the output terminal. Therefore, the periodic oscillations in the input voltage create a periodic flow of current through the pMOS and nMOS transistors. To account for every zero crossing event in the pendulum, the counter observable is defined as the sum of integrated currents $N_{p/n}(t)$ through both transistors, i.e. $y(t)=N_{p}(t)+N_{n}(t)$. Hence, the dynamics of the integrated electric current mimics the motion of the hand in a clock, with stochastic increments based on the periodic zero-crossing of the oscillator. We describe below the thermodynamically consistent modeling of this circuit to later analyze the cost of precise time-keeping.\par

\subsection{RLC oscillator}

We first describe the stochastic dynamics of the RLC circuit. The current and voltage in an RLC circuit oscillate as a result of the energy exchange between the electric field in the capacitor and the magnetic field in the inductor. In the absence of noise, the resistance damps these oscillations and any initial voltage will eventually relax to zero (the deterministic dynamics of the RLC circuit is given in the Appendix.~\ref{apsec: deterministic}). However, the noise in the RLC circuit, due to the thermal agitation of charge carriers in the resistor, will cause stochastic voltage oscillations around the steady-state value. The voltage fluctuations in the resistor have the Johnson-Nyquist form, i.e., its variance is proportional to the temperature $T$ \cite{johnson1928thermal,nyquist1928thermal}. Modeling a noisy resistor is done by replacing it with an ideal resistor and a random voltage source connected in series \cite{freitas2020stochastic}. This random voltage source has zero mean $\langle \xi(t) \rangle=0$ and has a delta-correlated spectrum, i.e., $\langle \xi(t)\xi(t') \rangle= 2k_B T R \,\delta(t-t')$. Applying Kirchoff's voltage law, the stochastic evolution of $V_{\rm{in}}(t)$ in Fig. \ref{fig:Circuit}-(a) is given by the following Langevin equation:
\begin{eqnarray}
LC_{\rm{in}} \;\Ddot{V}_{\rm{in}}(t) + RC_{\rm{in}} \;\dot{V}_{\rm{in}}(t) + V_{\rm{in}}(t) =  \xi(t),
\label{eqn: Langevin_input}
\end{eqnarray}
where $C_{\rm{in}}= C_i + 2 C_g$ is the effective capacitance at the input node of the inverter and $C_g$ is the gate-bulk capacitance of the transistors (Fig.~\ref{fig:Circuit}-(b)).  From Eq.~\eqref{eqn: Langevin_input}, the dynamics is equivalent to that of a damped harmonic oscillator in a thermal bath, with natural frequency $\omega = 1/\sqrt{LC_{\rm{in}}}$ and damping rate $\gamma= R/L$ \cite{van1992stochastic, pietzonka2022classical}.
The analogous position and velocity are the input voltage $V_
\text{in}$ and the current through the RLC loop, respectively. Note that the dynamics of the RLC oscillator is not affected in any way by the state of the CMOS inverter. Also, since there are no voltage or current sources powering it, the RLC oscillator will attain a thermal equilibrium state. This implies that the input voltage $V_\text{in}$ follows the Gibbs distribution $P_{\rm{eq}}(V_{\rm{in}})\propto e^{-C_{\rm{in}}V_{\rm{in}}^2/2q_e V_T}$, where $C_{\rm{in}}V_{\rm{in}}^2/2$ is the energy associated to the effective capacitance $C_{\rm{in}}$. The variance of $V_\text{in}$ is then given as $\sigma^2_{V_{\rm{in}}}=v_e^{\rm{in}} V_T$, where $v_e^{\rm{in}} \equiv q_e/C_{\rm{in}}$ is defined as the elementary voltage change associated to adding or removing an electron from the effective capacitance $C_\text{in}$. 

\subsection{CMOS inverter}

We now turn to the modeling of the CMOS inverter. 
The only degree of freedom in the CMOS inverter is the output voltage $v$. We can alternatively work with the output charge $q$, which is related to the output voltage as $q=C_{\rm{out}}v-C_{\rm{b}}V_{\rm{b}}$, where $C_{\rm{out}}= C_{\rm{b}} + 2 C_o$ is the effective output capacitance ($C_o$ is the drain-source capacitance of the transistors). 
The output voltage $v$ will change in time due to conduction of charge through the MOS transistors.
At the deterministic level and in the subthreshold mode of operation, the average electric current through the pMOS transistor is given as \cite{wang2006sub}:
\begin{equation}
 I_{p}(v, V_{\text{in}}; \,V_{\text{dd}})=I_0e^{(V_{\text{dd}}-V_{\text{in}}-V_{\text{th}})/(nV_T)}(1-e^{-(V_{\text{dd}}-v)/V_T}),
 \label{eqn: IV}
\end{equation}
where $I_0, V_{\rm{th}}$ and $n$ are the parameters that characterize the transistor (specific current, threshold voltage, and slope factor, respectively). For ease of calculations, we will consider the case when the slope factor is $n=1$. For the symmetric powering used in Fig. \ref{fig:Circuit}-(a), the average current through the nMOS transistor is $I_{n}(v, V_{\text{in}}; \,V_{\text{dd}})=I_{p}(-v,-V_{\text{in}};\,V_{\text{dd}})$.  

However, conduction through the CMOS transistors is noisy rather than deterministic. 
In the subthreshold mode of operation, the thermal noise is of shot noise nature, i.e., its variance is proportional to the average current \cite{sarpeshkar1993white,landauer1993solid, cui2008measurement}. Recently, thermodynamically consistent models have been developed to account for thermal shot noise in nonlinear electronic circuits \cite{freitas2021stochastic,gao2021principles}. Here, we will employ the formalism developed in \cite{freitas2021stochastic}. In this formalism, the transistors are modeled as externally controlled conduction channels with some associated capacitances, as shown in Fig.~\ref{fig:Circuit}-(b). The conduction of excess charges through the channels is modeled as a bi-Poissonian process. Hence, for each transistor $\Hat{\rho} \in \{p,n\}$, we associate forward ($+$) and backward ($-$) Poisson rates $\lambda_{\pm}^{p/n}(q)$ (they also depend implicitly on $V_\text{in}$, $V_\text{dd}$ and $V_b$), which give the probability per unit time for a jump $q\to q\pm \Delta_\rho q_e$ to occur, and $\Delta_\rho= \pm 1$ indicates the addition (+1) or removal (-1) of charges in the process  $\rho \in \{+p, -p, +n, -n\}$. In the case of the CMOS inverter, the forward (+) direction for the p(n)MOS transistor adds (removes) excess charge from the output terminal, i.e. $\Delta_{\pm p} =\pm 1$ and $\Delta_{\pm n} =\mp 1$.   Since the evolution of the input voltage $V_\text{in}$ occurs independently of the inverter output $v$, explicit time-dependent Poisson rates $\lambda_{\pm}^{p/n}(q,t)\equiv \lambda_{\pm}^{p/n}(q; V_{\rm{in}}(t))$ can be obtained for a given trajectory $\{V_{\rm{in}}(t)\}$. As explained in
\cite{freitas2021stochastic}, the functional form of the Poisson rates is determined by the deterministic I-V curve of the transistor (Eq. \eqref{eqn: IV}) and the requirement of local detailed balance (LDB). The LDB condition imposes that the log ratio of forward and backward rates associated with a given device must be related to the entropy change in the environment during an elementary jump. For example, the pMOS transistor at any time satisfies the LDB condition:
\begin{eqnarray}
\log\left(\frac{\lambda_+^p(q,t)}{\lambda_-^p(q+q_e,t)}\right)=-\frac{\delta
Q^p_{q\to q+q_e}}{k_B T}, \label{eqn: lbdp} 
\end{eqnarray} 
where $\delta
Q^p_{q\to q+q_e}= \phi(q+q_e) - \phi(q) -q_e V_{\rm{dd}}$ is the associated dissipated heat in the pMOS and $\phi(q) = q^2/(2C_{\rm{out}})+q C_{\rm{b}}V_{\rm{b}}/C_{\rm{out}}+const.$ is the internal energy of the circuit. Hence, one obtains the following rates for the pMOS transistor:

\begin{eqnarray} 
\lambda_+^p(q,t) &\!=\!& (I_0/q_e) \: e^{(V_\text{dd} -
V_\text{in}(t) - V_\text{th})/V_T}\label{eqn: rates_p} \\
\lambda_-^p(q,t) &\!=\!& \lambda_+^p(q,t) \: e^{((q+C_\text{b}V_\text{b})/C_{\rm{out}}-v_e^{\rm{out}}/2)/V_T}
e^{-V_\text{dd}/V_T}, \nonumber
\end{eqnarray} 
and for the nMOS transistor:
\begin{eqnarray}
\lambda_+^n(q,t) &\!=\!& (I_0/q_e)\: e^{(V_\text{in}(t) +
V_\text{dd} - V_\text{th})/V_T} \label{eqn: rates_n}\\
\lambda_-^n(q,t) &\!=\!& \lambda_+^n(q,t) \:
e^{-((q+C_\text{b}V_\text{b})/C_{\rm{out}}+v_e^{\rm{out}}/2)/V_T}e^{-V_\text{dd}/V_T}.\nonumber 
\end{eqnarray}
In the previous equations, we have defined $v_e^{\rm{out}} \equiv q_e/C_{\rm{out}}$ as the elementary voltage change associated to the jump of a charge $q_e$. The factor $e^{-v_e/2V_T}$ takes into account the
charging effects and becomes relevant at small scales and/or at low temperatures \cite{devoret1990effect,wasshuber2001computational,tucker1992complementary}, as will be discussed in the following section. Note that Eqs. \eqref{eqn: rates_p} and \eqref{eqn: rates_n} also define 
a natural time scale for the inverter dynamics, $\tau_0=(q_e/I_0) \: e^{V_{\rm{th}}/V_T}.$ For the rest of the article, we will also consider that the jump dynamics in the CMOS inverter is much faster than the slowest relevant timescale in the RLC, i.e. $\tau_0 \ll \tau_{\rm{RLC}} = \min[ \pi\sqrt{L C_{\text{in}}}, L/R]$.

The stochastic evolution of the charges in the output conductor for a given input signal, i.e. $\{q(\tau); V_\text{in}(\tau)\}$ in some time interval $\tau\in [0,t]$, is modeled as a continuous-time Markov jump process with the previous time-dependent rates. Therefore, it is characterized by the sequence of jumps $\{\rho_k\}$ along with their time stamps $\{\tau_k\}$, where the index $k$ is over all jumps. All this dynamical information is encoded in the instantaneous trajectory current for a given process $\rho \in \{+p, -p, +n, -n\}$, defined as follows:
\begin{eqnarray}
 j_\rho(q,t)=\sum_{k}\delta[\rho,\rho_k]\delta[q,q_{t_k}]\delta(t-t_k),
\end{eqnarray}
where $q_t$ is the state immediately before the instant $t$ and $\delta[x,y]$ is the Kronecker delta function.
Applying the charge conservation at the output node, we obtain:
\begin{eqnarray}
    q(t) &=& q(0) + q_e \left( \mathcal{N}_{+p}(t) - \mathcal{N}_{-p}(t) - \mathcal{N}_{+n}(t) + \mathcal{N}_{-n}(t)\right) \nonumber\\
    &=& q(0) + q_e \sum_\rho \Delta_\rho \: \mathcal{N}_\rho(t),
    \label{eqn: Langevin_output}
\end{eqnarray}
where $\mathcal{N}_\rho(t)=\int_0^t d\tau\; \sum_q j_\rho(q,\tau)$ is the total number of jumps of a particular process $\rho$ up to time $t$.
The time-integrated current $N_{\Hat{\rho}}(t)$ through a transistor $\Hat{\rho} \in \{n,p\}$ can then be obtained as follows,  
\begin{eqnarray}
    N_{\Hat{\rho}}(t) = (1/q_e)\int_0^t I_{\Hat{\rho}}(\tau)d\tau = \mathcal{N}_{+\Hat{\rho}}(t) - \mathcal{N}_{-\Hat{\rho}}(t),
\end{eqnarray}
where $I_{\Hat{\rho}}(\tau) = q_e \sum_q [j_{+\Hat{\rho}}(q,\tau)-j_{-\Hat{\rho}}(q+q_e\Delta_{+\Hat{\rho}},\tau)]$ is the instantaneous electric current through the transistor $\Hat{\rho}$.
From Eq.~\eqref{eqn: Langevin_output}, note that the output charge $q(t) = q(0)+ q_e[N_p(t)-N_n(t)]$ is related to the difference in integrated currents through the pMOS and nMOS transistors.

\section{Clock analysis}
\label{sec: Analysis}

\subsection{Counter Dynamics}
The dynamical quantity that will mimic the motion of the hand of a clock is the sum of the integrated current through both the pMOS and nMOS transistors, given as,  
\begin{equation}
y(t) \equiv  N_{p}(t) + N_{n}(t).
\end{equation}
The conductivity of the MOS transistors is controlled by the input voltage $V_\text{in}(t)$. Since $V_\text{in}(t)$ represents the position of the pendulum, this control couples the motion of the hand with the dynamics of the pendulum, and hence mimics the escapement mechanism in our circuit. Note that from Eqs. \eqref{eqn: rates_p} and \eqref{eqn: rates_n}, the voltage $V_{\rm{in}}$ does not affect in any way the log-ratio of the rates involved in the LDB condition, Eq. \eqref{eqn: lbdp}. Thus, the voltage oscillations at the input only affect the kinetic part of the rates. An essential difference between our circuit and the toy model considered in \cite{pietzonka2022classical} is that in our case the rates depend on the oscillatory input in a continuous way, while in the toy model the rates were only a function of the sign of the input. Those coarse-grained rates used in the toy model can be recovered in our circuit by considering the single-electron regime of operation, which is discussed below.

\subsection{Thermodynamic efficiency of the clock}

For fixed power $V_{\rm{dd}}$, the circuit relaxes to a non-equilibrium steady state. The total steady-state rate of dissipation of the circuit can be divided into the contributions from the RLC and the CMOS inverter, $T\dot{\sigma} = T\dot{\sigma}_{\rm{RLC}} + T\dot{\sigma}_{\rm{Inv}} $. Since the RLC part is at thermal equilibrium, its dynamics is free of any thermodynamic cost, i.e. $\dot{\sigma}_{\rm{RLC}}=0$. Therefore, the dissipation of the circuit is due only to that of the CMOS inverter, $\dot{\sigma} = \dot{\sigma}_{\rm{Inv}}$. At steady state, the total dissipation up to time $t$ is equal to the work done by the powering sources, given as
\begin{equation}
\begin{aligned}
    T\dot{\sigma}t \;&=  q_e\langle N_{p}(t)\rangle \, \Delta V\\
    &= q_e\langle y(t) \rangle \, (\Delta V/2)
\end{aligned}
\end{equation}
In the last line, we used the equality of the average steady currents through the transistors, $\langle N_{p}(t)\rangle = \langle N_{n}(t)\rangle$. This is the result of charge conservation in the output node $v(t)$, which can be seen by taking the average in Eq.~\eqref{eqn: Langevin_output}. 

As any clock, this circuit is a thermal machine that produces entropy to measure the passage of time. To quantify its thermodynamic efficiency, we will study the product between the relative uncertainty in the counting process $\text{Var}[y(t)]/\langle y(t) \rangle^2$ and the associated entropy production $\dot{\sigma}\,t$. This uncertainty product $\mathcal{Q}$, is hence defined as
\begin{eqnarray}
    \mathcal{Q}\equiv\lim_{t\to\infty}\frac{{\rm Var}[y(t)]}{\langle y(t) \rangle^2} \: \frac{\dot{\sigma} t}{k_b}.
 \end{eqnarray}
For overdamped systems, the above quantity is bounded from below by the TUR $\mathcal{Q} \geq 2$ \cite{horowitz2020thermodynamic}. This bound then imposes a minimum thermodynamic cost in order to obtain precise currents in a system. As in \cite{pietzonka2022classical}, we want to study the behaviour of $\mathcal{Q}$ for systems that have an underdamped component, such as the RLC oscillator in our circuit. 

 \begin{figure*}[ht!]
    \centering
    \hspace{-1.3cm}
    \subfloat[]{{\includegraphics[trim=0 250 0 0, clip,width=0.55\textwidth]{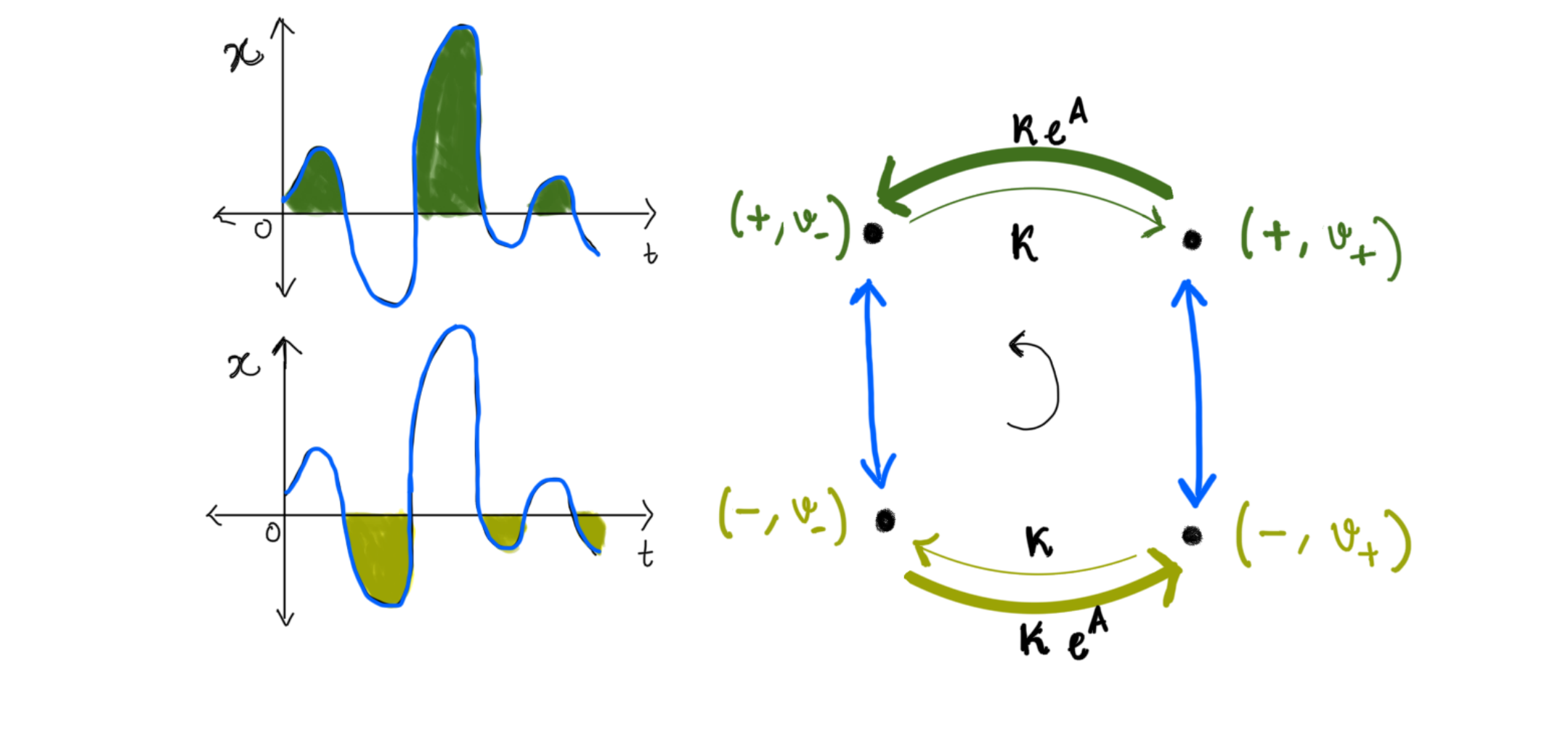} }}%
    \subfloat[]{{\includegraphics[trim=0 0 0 50, clip,width=0.5\textwidth]{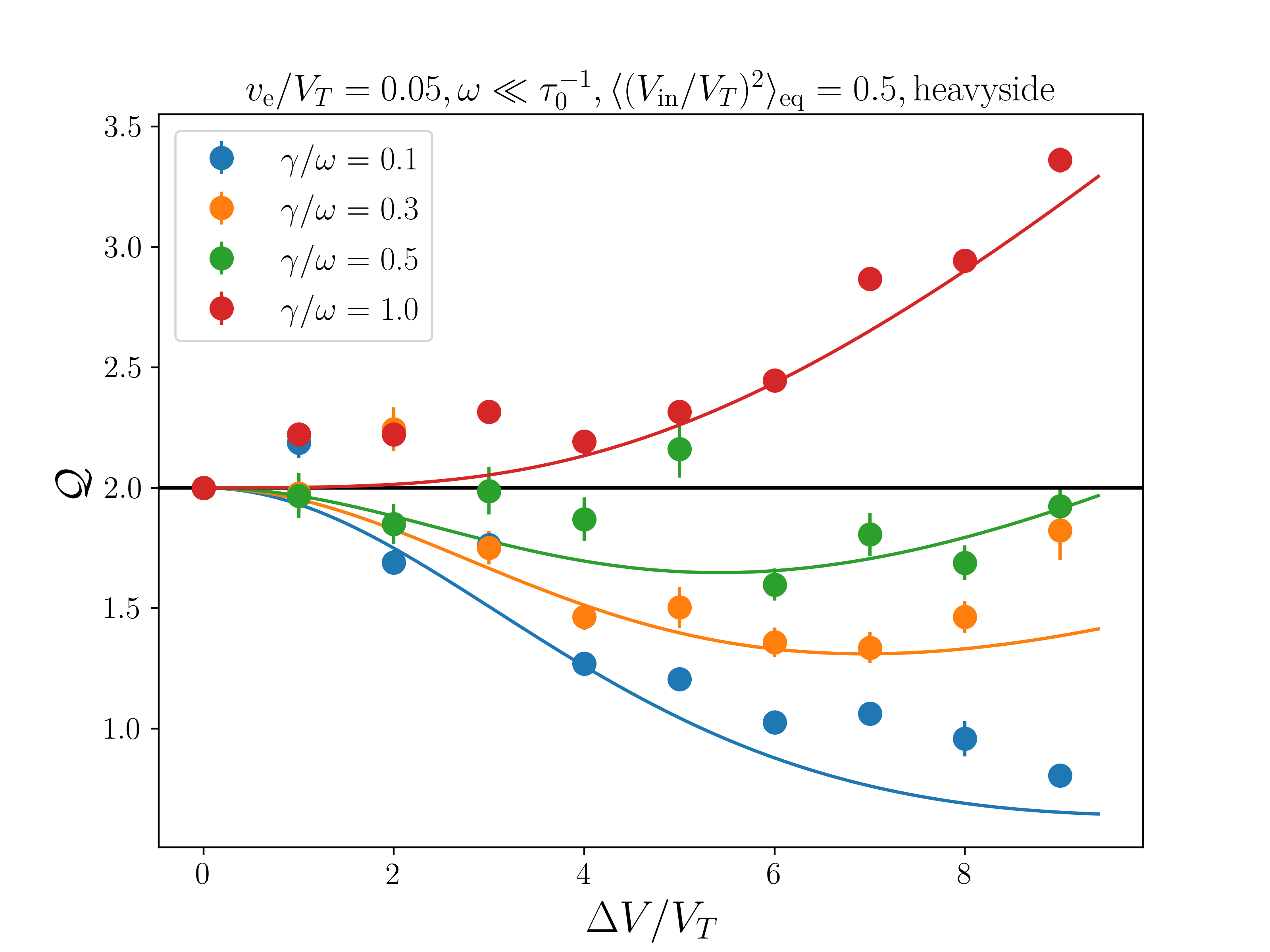} }}%
    \caption{(a): An equivalent 4-state representation of the circuit in the single-electron regime, to compare with the toy model \cite{pietzonka2022classical}. In this representation, $\{+,-\}$ are the coarse-grained states of the input voltage when $V_{\rm{in}}(t)\ge0$ and $V_{\rm{in}}(t)<0$ respectively, and $\{v_+,v_-\}$ are the states of the output voltage $v(t)$. Here, $k=(1/\tau_0)e^{V_{\rm{in}}^{\rm{th}}/V_T}$ is a kinetic constant defined on the basis of a threshold input voltage $V_{\rm{in}}^{\rm{th}}$, and $A\equiv V_{\rm{dd}}/V_T$ is the thermodynamic affinity of the cycle. The counting variable $y(t) = N_\rightarrow(t|-) + N_\leftarrow(t|+), $ is obtained using the conditional integrated current $N_\rightarrow(t|-)$ and $N_\leftarrow(t|+)$ through the lower ($-$) and upper ($+$) branches, respectively. (b): The uncertainty product $\mathcal{Q}$ of the counter $y(t)$ in the single-electron regime. It is plotted as a function of the voltage difference $\Delta V/V_T = 2V_{\rm{dd}}/V_T$ for different damping rates of the RLC circuit. The markers are obtained from Gillespie circuit simulations, and solid curves are obtained using Eq.~\eqref{eqn: Q_toymodel}. Parameters: $\omega \equiv 1/\sqrt{LC_{\rm{in}}} = 0.1\tau_0^{-1}, V_{\rm{in}}^{\rm{th}}/V_T=10, v_e^{\rm{out}}/V_T = 10$ and $C_bV_b = -q_e/2$.}
    \label{fig: 2state_heaviside}%
\end{figure*}

\subsection{Single-electron regime}
\label{sec: toymodel}

Electronic circuits can be fabricated on different scales. Typically, capacitances and inductances scale linearly with the characteristic length of the components \cite{gopal2022large}. Thus, the elementary voltage changes $v^\text{in/out}_e = q_e/C_\text{in/out}$ defined above will increase as the circuit is scaled down. Modern-day CMOS transistors (sub-$7nm$ fabrication processes) have associated capacitances as low as $C\simeq 10aF$ \cite{Zheng:EECS-2016-189,hofheinz2006coulomb}, which implies that the elementary voltage change is $v_e \simeq 16 mV$. This is comparable to the thermal voltage $V_T =26mV$ at room temperature. The circuit could also be operated at low temperatures, and for example values as high as $v_e/V_T\simeq 50$ can be achieved at $T\simeq 4K$. Such nanoscopic devices at low temperatures are the working ground for single-electron devices, as the charging energy to move an elementary charge scales as $q_e^2/2C = (1/2)C v_e^2$. Therefore, the energy levels associated with discrete numbers of excess charges become well separated, which allows precise control over few charges or even single electrons/holes, as in the case of a single-electron quantum dot \cite{wasshuber2001computational, kastner1992single, fallahi2005imaging}. In the following, we will analyze separately the limits $v_e^\text{in}/V_T \gg 1$ and $v_e^\text{out}/V_T \gg 1$ for our clock circuit.

Firstly, in the limit $v_e^{\rm{in}}/V_T\gg1$ the dynamics of the counter $y(t)$ converges to an auxiliary dynamics with coarse-grained rates, which only depend on the sign of $V_\text{in}$.
This coarse-grained escapement coupling is equivalent to the one discussed in the toy model \cite{pietzonka2022classical}.
To see this, we first note that the transition rates satisfy $\lambda_\pm^n \propto e^{V_\text{in}(t)/V_T}$ and $\lambda_\pm^p \propto e^{-V_\text{in}(t)/V_T}$. Thus, whenever $|V_\text{in}(t)/V_T| \gg 1$, only one of the transistors is effectively active, and we can disregard the presence of the other. Secondly, we note that for $v_e^\text{in}/V_T \gg 1$ the previous condition $|V_\text{in}(t)/V_T| \gg 1$ is satisfied for almost the entire period of the input signal, with the exception of brief intervals of duration $\delta \tau$ around the zero crossings. This is due to the fact that the amplitude of stochastic oscillations increases with $v_e^\text{in}$ (note that the variance of $V_\text{in}$ is given by $\sigma^2_{V_\text{in}} = v_e^\text{in} V_T$). Thus, if $\delta\tau \ll \tau_0 \ll \tau_{\rm{RLC}}$, where $\tau_0$ and $\tau_{\rm{RLC}}$ are the timescales associated with the counter and the RLC respectively, the counter rates will almost always be such that only one transistor is effectively active, depending only on the sign of $V_\text{in}(t)$. The condition $\delta\tau \ll \tau_0$ can always be achieved by increasing $v_e^\text{in}/V_T$ (see Appendix.~\ref{apsec: Coarse-grained-coupling}). Also, since $\tau_0 \ll \tau_{\rm{RLC}}$, the counter dynamics will quickly reach equilibrium with the corresponding voltage source during each half-period of oscillation. Once equilibrium is reached, the forward and backward jumps of the active transistor do not contribute to the net progress of the counter variable $y(t)$, regardless of the magnitude of the input voltage $|V_\text{in}(t)/V_T|$. Thus, the dependence of the rates on the input voltage can be simplified to $\lambda_{\pm}^n \propto \Theta(V_\text{in}/V_T)$ and $\lambda_{\pm}^p \propto \Theta(-V_\text{in}/V_T)$, where $\Theta(x)$ is a Heaviside function such that $\Theta(x) = 1$ for $x\ge0$ and $0$ elsewhere.
In such an auxiliary dynamics, the counter $y(t)$ progresses only when the input voltage changes its sign due to relaxation of the output voltage from one equilibrium to the other.

Secondly, in the limit $v_e^{\rm{out}}/V_T\gg1$, the dynamics of the output voltage $v$ can be restricted to a few states. By additionally controlling the biasing voltage $V_b$, the state space of the charges in the output conductor ($q$) can even be restricted to only two degenerate states. This can be clearly seen from the equilibrium ($V_{\rm{dd}} =0$) distribution $P_{\rm{eq}}(q)\propto e^{-(v_e^{\rm{out}}/2q_e^2V_T)[q+ C_b V_b]^2}$. For $v_e^{\rm{out}}/V_T\gg1$, the probability distribution is sharply peaked around the states near the minimum of the internal energy of the circuit, $\phi(q) \propto v_e^{\rm{out}}[q+ C_b V_b]^2$.  When the output bias is fixed such that $C_b V_b = - (j-1/2) q_e$ where $j$ is an integer, then the state space for $q$ is restricted to $\{(j-1) q_e,\, j q_e\}$, and at equilibrium both are equiprobable. The corresponding voltage values are $v_- \equiv -v_e^{\rm{out}}/2$ and $v_+ \equiv v_e^{\rm{out}}/2$. For any potential difference $\Delta V < v_e^{\rm{out}}$, the state space can still be restricted to two states due to the high charging energy, but the steady-state distribution can be biased towards $q=(j-1) q_e$ or $q=j q_e$ depending on the biasing voltage. \par
 
\section{Results}
\label{sec: Results}
In this section, we will explore the behaviour of the circuit in different regimes of operation. First, we will look at the complete single-electron regime, $v_e^{\rm{in}}/V_T\gg1$ and $v_e^{\rm{out}}/V_T\gg1$, where the dynamics of the counter $y(t)$ converges to that of the toy model and the TUR can be violated. After that, we will sequentially relax the conditions $v_e^{\rm{in}}/V_T\gg1$ and $v_e^{\rm{out}}/V_T\gg1$.

\subsection{Single-electron regime: Convergence with the toy model} 


As explained above, in the full single-electron regime, the output voltage can be restricted to only two states $v_\pm$. In addition, transitions between these two states occur at rates that depend only on the sign of the input voltage. Explicitly, the rates read
\begin{flalign}
    k_\rightarrow^p &\equiv \lambda_+^{p}((n-1) q_e) = k\,\Theta(-V_{\rm{in}}(t)/V_T)e^{V_{\rm{dd}}/V_T}\\
    k_\leftarrow^p &\equiv\lambda_-^{p}(n q_e) = k\,\Theta(-V_{\rm{in}}(t)/V_T) \nonumber\\
    k_\leftarrow^n&\equiv \lambda_+^{n}(n q_e) = k\,\Theta(V_{\rm{in}}(t)/V_T) e^{V_{\rm{dd}}/V_T}\\
    k_\rightarrow^n&\equiv\lambda_-^{n}((n-1) q_e) = k\,\Theta(V_{\rm{in}}(t)/V_T) ,\nonumber  
\end{flalign}
where $k$ is a kinetic constant. In Fig.~\ref{fig: 2state_heaviside}-(a), we give a minimal 4-state representation of the dynamics of the circuit in the single-electron regime. Since the rates are coarse-grained, the input voltage $V_{\rm{in}}(\tau)$ can also be coarse-grained to only 2 states $x\equiv\{+,-\}$ based on $V_{\rm{in}}\ge0$ and $V_{\rm{in}}<0$ respectively. In this representation, the vertical transitions $(\pm\to\mp)$ correspond to changes in the coarse-grained state of the RLC, and we interpret these zero-crossing transitions as ticks in the pendulum. Similarly, the horizontal transitions $(v_{\pm}\to v_{\mp})$ correspond to the transitions in the pMOS (-) and nMOS (+) transistors, based on the state of the RLC. The counting observable $y(t)$ can be equivalently computed by the sum of the integrated currents through both horizontal branches as
\begin{eqnarray}
    y(t) = N_\rightarrow(t|-) + N_\leftarrow(t|+), 
\end{eqnarray}
where we used the equality $N_{p}(t) = N_\rightarrow(t|-)$ and $N_{n}(t)=N_\leftarrow(t|+)$ due to the coarse-grained rates. For finite voltage differences $V_{\rm{dd}}\neq 0$, the most probable trajectory will be due to the following cycle $(-,v_-)\to(-,v_+)\to(+,v_+)\to(+,v_-)\to(-,v_-)$. In this cycle, the counter $y(t)$ has unit increments for each tick event $(\pm\to\mp)$. This is exactly the counter dynamics of the toy model \cite{pietzonka2022classical} which is also defined in the same state space $\{v_-,v_+\}$ and with the same coarse-grained rates $k(y_{\pm}\to y_{\mp})$. The pendulum trajectories in the toy model are also given independently by an underdamped harmonic oscillator, as defined in Eq.~\eqref{eqn: Langevin_input}. The powering voltage in the inverter plays the role of the thermodynamic affinity in the toy model, i.e. $A\equiv \Delta V/2V_T = V_{\rm{dd}}/V_T.$  Therefore, the stochastic dynamics of the counter $y(t)$ in the circuit converges to that of the toy model in the single-electron regime. \par

Using the time-scale separation between the CMOS inverter and the RLC, $\tau_0\ll 2\pi/\omega$, we can analytically derive the uncertainty product $\mathcal{Q}$ with coarse-grained rates (see Section.~\ref{sec: coarse_grained_derivation}), for any arbitrary number of states in the output voltage. In the 2-state limit, we reobtain the same expression as the toy model \cite{pietzonka2022classical}, given by 
\begin{equation}
    \mathcal{Q} \!= \!\frac{2V_{\rm{dd}}}{V_T}\!\left[\frac{1}{\sinh(V_{\rm{dd}}/V_T)}+ \frac{D_N}{\langle \dot{N} \rangle}\tanh\left(V_{\rm{dd}}/2V_T\right)\right],
    \label{eqn: Q_toymodel}
\end{equation}
where $D_N/\langle \dot{N} \rangle$ is the relative uncertainty of the number of ticks $N(t)$ in the RLC up to time $t$. This quantity is an increasing function of the damping $\gamma$ in the RLC, and can be obtained analytically as shown in \cite{pietzonka2022classical} (See Appendix.~\ref{apsec: precision_ticks}).

In Fig.~\ref{fig: 2state_heaviside}, we plot the uncertainty product $\mathcal{Q}$ of the circuit from the numerical simulations with the coarse-grained rates for different damping rates $\gamma$. The numerical results show a violation of the TUR bound when the RLC has low damping, in agreement with the expression above (Eq.~\eqref{eqn: Q_toymodel}). 
The critical value of damping below which there can be violations of the TUR is for $D_N/\langle \dot{N} \rangle<1/3$, which corresponds to $\gamma/\omega<0.981$, as given in \cite{pietzonka2022classical}. With a lower damping in the RLC, the ticks are more coherent and, when combined with a higher thermodynamic affinity ($\Delta V$), the counting process also becomes more precise (Fig.~\ref{fig: app_2state_coarse}. (a)). The global minima of the uncertainty product of the clock vanishes, i.e. $\mathcal{Q}_{\rm{min}}\to 0$, only for $D_N/\langle \dot{N} \rangle\to0$ and $V_{\rm{dd}}\to\infty$. Both of these requirements are not practical, as they require a vanishing resistance in the RLC and an infinite voltage supply in the inverter. 

Both restrictions $v_e^{\rm{in}}/V_T\gg1$ and $v_e^{\rm{out}}/V_T\gg1$ play a crucial role in the violation of the TUR. Firstly, through the coarse-grained coupling, the stochastic dynamics of the counter state $y(t)$ becomes strongly dependent on the periodicity of the pendulum and counts only the number of ticks in the pendulum, $N(t)$. Secondly, in the two-state limit, the counting process given a tick is also precise due to the restricted state space of the output voltage $v(t)$. In the following sections, we consider the other regimes of operation by relaxing these constraints and compare the uncertainty product with the TUR bound.\par

\subsection{Role of 2-state limit}

To study the role of the 2-state dynamics we will first relax the constraint at the output node by considering $v_e^{\rm{out}}/V_T<1$ so that the output voltage is no longer restricted to two states, but with $v_e^{\rm{in}}/V_T\gg1$ such that the coupling can still be coarse-grained.  
This regime can be implemented by simply increasing the bias capacitance. Given a coarse-grained state ($\pm)$ at the input node and again assuming the timescale separation, the output voltage $v(t)$ relaxes to an equilibrium distribution peaked around the powering voltage ($\mp V_{\rm{dd}}$) of the conductive transistor (n/pMOS). This implies that for every tick event $(\pm\to\mp)$, the output voltage must change between these two distributions, and the counter $y(t)$ increments by $\mathcal{O}(\Delta V/v_e^{\rm{out}})$, where $\Delta V = 2 V_{\rm{dd}}$ is the applied voltage difference.  

\begin{figure}[h!]
    \centering
    \includegraphics[trim=0 0 0 50, clip,width=0.5\textwidth]{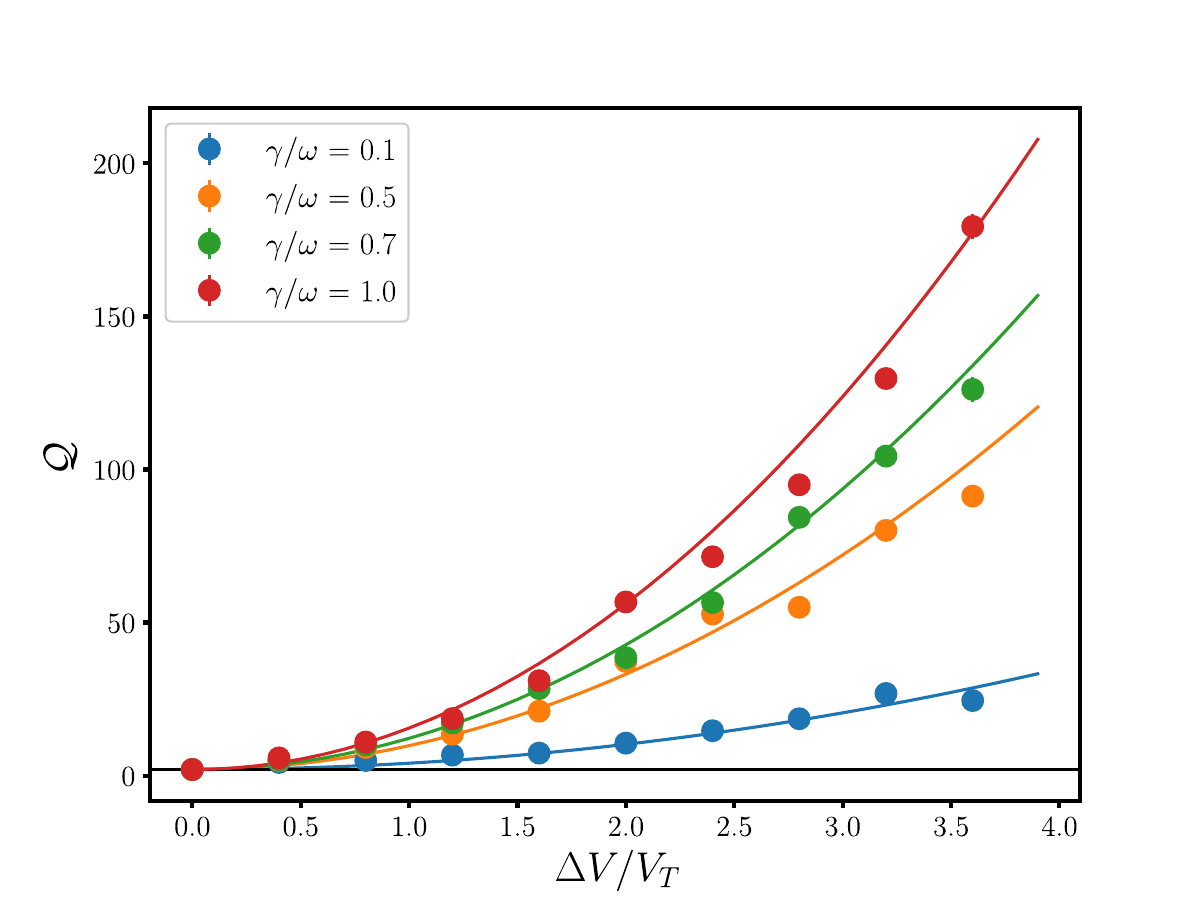}
    \caption{The uncertainty product $\mathcal{Q}$ of the counter $y(t)$ with coarse-grained coupling and a macroscopic state space. It is plotted as a function of the voltage difference $\Delta V/V_T = 2V_{\rm{dd}}/V_T$ for different damping rates. The markers are obtained from the Gillespie simulations of the circuit and the solid curves are obtained using Eq.~\eqref{eqn: Q_macroscopic}. Parameters: $\omega \equiv 1/\sqrt{LC_{\rm{in}}} = 0.1\tau_0^{-1}, V_{\rm{in}}^{\rm{th}}/V_T=10, v_e^{\rm{out}}/V_T = 0.05$ and $C_bV_b = -q_e/2$.}
    \label{fig: macroscopic_heaviside}
\end{figure}

Using the generalized expression for the uncertainty product with coarse-grained rates (see Sec.~\ref{sec: coarse_grained_derivation}), we derive $\mathcal{Q}$ for a macroscopic state space at the inverter output (i.e., $v_e^{\rm{out}}/V_T\to0$) \cite{gopal2022large},
\begin{eqnarray}
        \mathcal{Q} = 2\left[1+ \frac{D_N}{\langle \dot{N} \rangle}\frac{V_{\rm{dd}}^2}{v_e^{\rm{out}}V_T}\right].
        \label{eqn: Q_macroscopic}
\end{eqnarray}
As shown in Fig.~\ref{fig: macroscopic_heaviside}, the TUR bound is 
restored for any damping of the RLC. This is also obvious from
Eq.~\eqref{eqn: Q_macroscopic}, as the relative uncertainty of the ticks is always positive, $D_N/\langle \dot{N} \rangle>0$. The restoration of the TUR can be qualitatively understood because of the higher uncertainty in the counting process with a larger state space. Compared to the unit increments in the 2-state regime, there is a larger number of counter transitions (of the order of $\Delta V /v_e^{\rm{out}}$) for each tick in the RLC. 
We note that Eq.~\eqref{eqn: Q_macroscopic} gives an accurate prediction for the uncertainty product $\mathcal{Q}$ even for 
a limited counter state space ($V_{\rm{dd}}/v_e^{\rm{out}}\sim\mathcal{O}(10)$ in Fig.~\ref{fig: macroscopic_heaviside}). This implies that the violation of the TUR is not possible for a mesoscopic counter.

As discussed in more detail in Sec.~\ref{sec: coarse_grained_derivation}, it is possible to see that in this case the entropy production rate increases with the voltage difference as $\Dot{\sigma} \propto \Delta V^2/v_e^{\rm{out}}V_T$ (see also Fig.~\ref{fig: app_macroscopic_coarse}). In contrast, in the 2-state limit one finds the linear growth $\Dot{\sigma} \propto \Delta V /V_T$, for $\Delta V/V_T \gg1$ (Fig.~\ref{fig: app_2state_coarse}).  Therefore, when $v_e^{\rm{out}}/V_T\ll1$ (macroscopic state space), the counter produces more entropy to achieve a similar precision, resulting in $\mathcal{Q}$ being orders of magnitude larger compared to the 2-state limit.

\subsection{Role of the coarse-grained coupling}
We will now explore the role of coarse-grained coupling in the violation of the TUR bound. We will thus consider $v_e^{\rm{in}}/V_T<1$ at the input node but still with $v_e^{\rm{out}}/V_T\gg1$ so that the counter-dynamics can be restricted to 2 states. Then, the transition rates can no longer be coarse-grained. This corresponds to a circuit with a macroscopic RLC oscillator, while still keeping a nanoscale CMOS inverter. Since the input voltage fluctuations scale with $v_e^{\rm{in}}/V_T$, the counter dynamics $y(t)$ will explicitly depend on the value of $V_{\rm{in}}(t)$ for the small-amplitude oscillations in this regime, and not only on its sign. Therefore, the dynamics of the circuit around $V_{\rm{in}}(t)\simeq 0$ will play an important role, which is lacking in the toy model \cite{pietzonka2022classical}.\par

\begin{figure}[h!]
    \centering
    \includegraphics[trim=0 0 0 40, clip,width=0.5\textwidth]{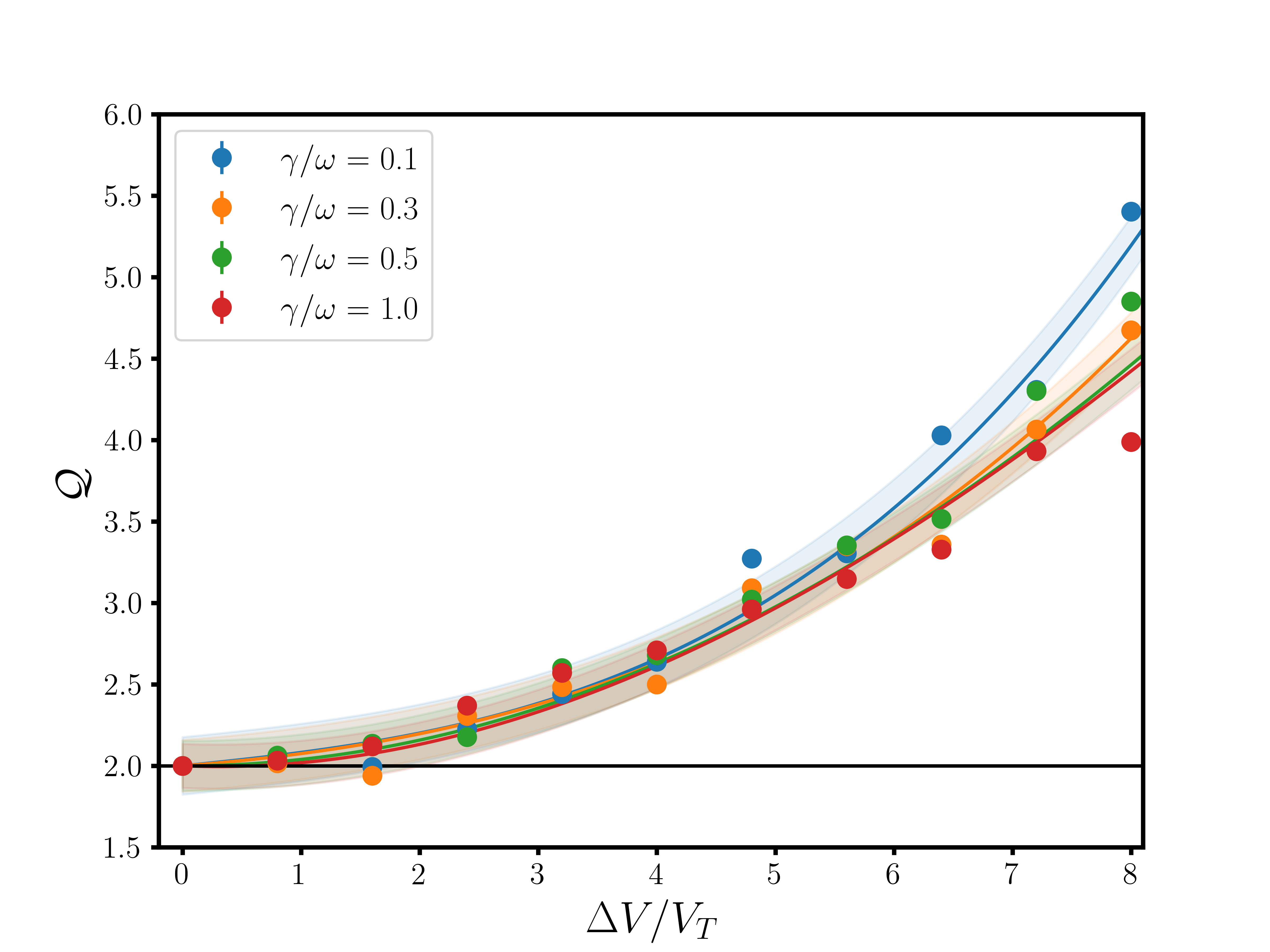}
    \caption{The uncertainty product $\mathcal{Q}$ of the counter $y(t)$ in the 2-state limit but without the coarse-grained coupling. It is plotted as a function of the voltage difference $\Delta V$ for different damping rates. The markers are obtained from the Gillespie simulations of the circuit and the solid curves are obtained using the approximate dynamics of Eq.~\eqref{apeqn: Gaussian_timescale}. Parameters: $\omega \equiv 1/\sqrt{LC_{\rm{in}}} = 0.1\tau_0^{-1}$, $v_e^{\rm{in}}/V_T = 0.1$, $v_e^{\rm{out}}/V_T = 10$ and $C_bV_b = -q_e/2$.}
    \label{fig: 2_state_true}
\end{figure}

As seen in Fig.~\ref{fig: 2_state_true}, the TUR bound is again restored if the rates cannot be coarse-grained, even in the two-state limit.  This is due to the larger amount of entropy production in the counter dynamics when $V_{\rm{in}}\simeq 0$, where both devices are equally conductive. Since the jump timescale $\tau_0$ is smaller than or comparable to $\delta t$ in this regime, the counter will have increments near $V_{\rm{in}}=0$ in addition to the unit increment with every tick. These undesirable increments correspond to a current flow through both transistors near $V_{\rm{in}}=0$, which can be neglected for the coarse-grained coupling.  In Fig.~\ref{fig: 2_state_true}, we also find that an approximate Langevin dynamics for the counter $y(t)$ (Appendix.~\ref{apsec: Approximate counter dynmaics}) captures the uncertainty product $\mathcal{Q}$ computed using exact stochastic simulations of the circuit. 
\par

\subsection{Macroscopic regime of the full circuit}

\begin{figure}[h!]
    \centering
    \includegraphics[trim=0 0 0 40, clip,width=0.5\textwidth]{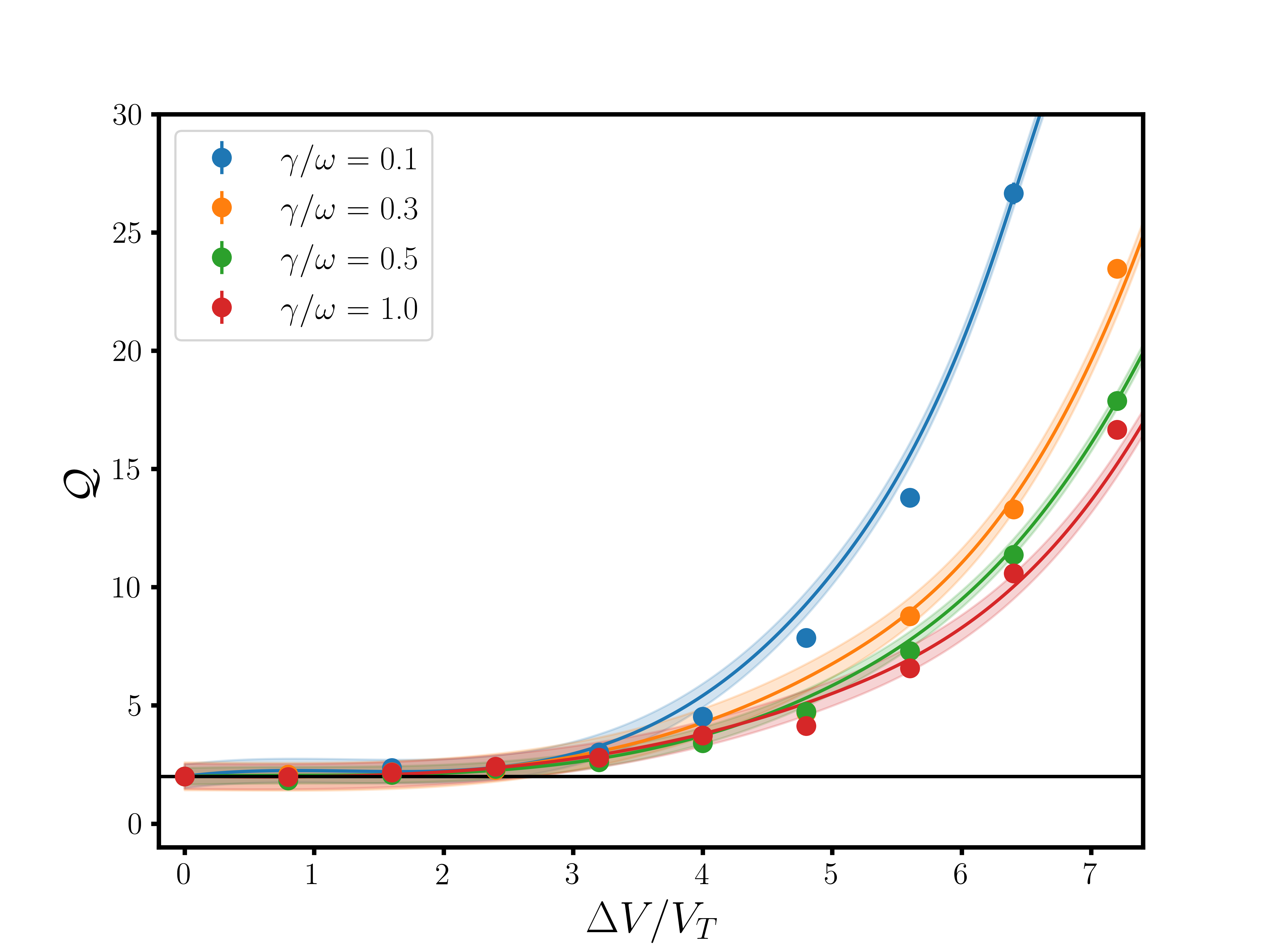}
    \caption{The uncertainty product $\mathcal{Q}$ of the counter $y(t)$ in the macroscopic operating regimes. It is plotted as a function of the voltage difference $\Delta V$ for different damping rates of the RLC circuit. The markers are obtained from the Gillespie simulations with the time-dependent rates of the circuit. The solid curve and the highlighted band represent the interpolated curves and the standard error computed using the approximate dynamics of Eq.~\eqref{apeqn: Gaussian_timescale}.  Parameters: $\omega \equiv 1/\sqrt{LC_{\rm{in}}} = 0.1$, $v_e^{\rm{in}}/V_T = 0.1$, $v_e^{\rm{out}}/V_T = 0.1$ and $C_bV_b = -q_e/2$. }
    \label{fig: macroscopic_true}
\end{figure}

Finally, relaxing both constraints on the elementary voltages such that $v_e^{\rm{out}}/V_T<1$ and $v_e^{\rm{in}}/V_T<1$, we consider the macroscopic regime of the full circuit. In this regime, the RLC will have stochastic oscillations of small amplitude that will be coupled to a counter with large increments per tick ($\mathcal{O}(V_{\rm{dd}}/v_e^{\rm{out}})$). In Fig.~\ref{fig: macroscopic_true}, we plot the uncertainty product ($\mathcal{Q}$) of the clock for different damping rates of the RLC circuit. As expected from the previous sections, the circuit does not violate TUR for any damping. In this regime, there is a larger production of entropy in the counter due to the current flow around $V_{\rm{in}}\eqsim 0$ and also due to the macroscopic state space of the counter. Hence, we find that in the macroscopic limit of the circuit, the TUR bound still provides a minimum thermodynamic cost needed for precise currents.
As seen in Fig.~\ref{fig: macroscopic_true}, the approximate Langevin dynamics for the counter $y(t)$ (Appendix.~\ref{apsec: Approximate counter dynmaics}) captures the behavior of the uncertainty product $\mathcal{Q}$, and compares well with the exact stochastic simulations of the circuit.

Finally, we note that while the uncertainty product $\mathcal{Q}$ decreases with decreasing damping factor for coarse-grained rates (Figures \ref{fig: 2state_heaviside}-(b) and \ref{fig: macroscopic_heaviside}), it has the opposite behaviour in the other regimes (Figures \ref{fig: 2_state_true} and \ref{fig: macroscopic_true}). As we show in Appendix \ref{ap: spliting}, the dependence of $\mathcal{Q}$ on the damping factor $\gamma/\omega$ is due only to the variance in the counting observable $\text{Var}[y(t)]$. 
The unexpected behaviour of $\mathcal{Q}$ in Figures \ref{fig: 2_state_true} and \ref{fig: macroscopic_true} (the fact that it decreases with increasing damping) is related to an interesting feature of the current fluctuations in the CMOS inverter first identified in \cite{gopal2022large}.
There, it was shown that, above a certain value of powering voltage, the variance of the steady-state current through the inverter has a local minimum at $V_\text{in} = 0$. This variance enters as an input in the coupled Langevin model of Appendix \ref{apsec: Approximate counter dynmaics} used to reproduce the numerical results, which explains why increasing the damping (and therefore the amount of time the input signal spends in the neighborhood of $V_\text{in} = 0$) decreases the variance of the current through the inverter and therefore that of the counter variable $y(t)$. 

\section{Uncertainty product with coarse-grained coupling}
\label{sec: coarse_grained_derivation}
In this section, we generalize the derivation of the uncertainty product $\mathcal{Q}$ done in the toy model \cite{pietzonka2022classical} to this circuit with coarse-grained rates ($v_e^{\rm{in}}/V_T\gg1$), but for any arbitrary number of states in the output voltage. With coarse-grained rates, the entire dynamics of the circuit depends only on the coarse-grained states of the input voltage $\{+,-\}$. 
Since we assume a time-scale separation between the RLC and the CMOS inverter, i.e.  $\tau_0/ \tau_{\rm{RLC}}\ll 1$, the output voltage $v(t)$ also relaxes quickly to the equilibrium distribution $P_{\rm{eq}}^{+/-}(v)$ corresponding to the source of the conductive transistor. When the input state is $+(-)$, the conductive transistor is the n(p)MOS transistor and the output voltage relaxes to $-(+)V_{\rm{dd}}$. Hence, the equilibrium distributions corresponding to the coarse-grained states $+$ and $-$ of the RLC are given as,
\begin{eqnarray}
    P_{\rm{eq}}^{+} (v) &\propto& e^{-\frac{1}{(2v_e^{\rm{out}} V_T)}\left[v +V_{\rm{dd}}\right]^2}\label{eqn: Peq_+}\\
    P_{\rm{eq}}^{-} (v) &\propto& e^{-\frac{1}{(2v_e^{\rm{out}} V_T)}\left[v - V_{\rm{dd}}\right]^2}.\label{eqn: Peq_-}
\end{eqnarray} 
The counter state $y(t)$ changes only when there is a tick and the stationary distribution relaxes to the new equilibrium distribution. The counter state $y(t)$ can then be described as the sum of independent increments $\Delta y_{i}$, given as
\begin{eqnarray}
    y(t) &=& N_{p}(t) + N_{n}(t)\\
    &=& \sum_{i=0}^{N(t)-1} \Delta y_{i},
\end{eqnarray}
where $N(t)=\int_0^td\tau|\dot{V}_{\rm{in}}(\tau)|\delta[V_{\rm{in}}(\tau)]$ is the number of ticks until time t and $\Delta y_{i}$ is the change in $y$ between $i^{th}$ and $(i+1)^{th}$ ticks.  Since only one of the transistors is conductive between 2 ticks, all the changes in the output voltage are due to that conductive transistor (Eq.~\eqref{eqn: Langevin_output}). Hence, $\Delta y_{i}$ can be computed as follows,
\begin{eqnarray}
    \Delta y_{i} = s(i)\frac{(v_{i}-v_{i+1})}{v^{\rm{out}}_e}
\end{eqnarray}
where $v_{i}$ is the output voltage before the $i^{th}$ tick and $s(i)\equiv \{+,-\}$ is the coarse-grained state between the $i^{th}$ and the $(i+1)-{th}$ tick.\par
Using the equilibrium distributions (Eqs.~\eqref{eqn: Peq_+},~\eqref{eqn: Peq_-}), we can compute the mean and dispersion of $\Delta y$, given as
\begin{eqnarray}
    \langle \Delta y \rangle &=& \frac{\langle v \rangle_{-}-\langle v \rangle_{+}}{v^{\rm{out}}_e},\\
    \sigma^2(\Delta y) &=& \frac{(\sigma^2_{-}(v)+\sigma^2_{+}(v))}{{v^{\rm{out}}_e}^2},
\end{eqnarray}
where we also used the independence of the voltages before two consecutive ticks. For a given RLC trajectory up to time $t\gg1/\omega$ with $N(t)\gg1$ ticks, the probability of the counter state $P(y|N)$ obeys the central limit theorem with mean $\langle y|N\rangle = N \langle \Delta y \rangle$ and variance $\sigma^2(y|N) = 2 N \sigma^2(\Delta y)$. The factor 2 in the variance is the result of the dependence between any two consecutive increments $\Delta y_i$ \cite{2problem}. This characterization is possible since the RLC dynamics is independent of changes in the CMOS inverter. Similarly, one can also assume the central limit theorem for the statistics of $N(t)$, such that $P(N,t)$ is also a Gaussian with mean $\langle \dot{N} \rangle t$ and variance $2 D_N t$. As the RLC dynamics is just that of the underdamped harmonic oscillator, both the mean and the dispersion of ticks can be obtained semi-analytically (Appendix.~\ref{apsec: precision_ticks}),
\begin{eqnarray}
     \langle \dot{N} \rangle &=&\omega/\pi\nonumber\\
     D_N &=& \frac{\langle \dot{N} \rangle}{2}+\int_{0^+}^{\infty}d\tau\left[\langle \dot{N}(0)\dot{N}(\tau)\rangle - \langle \dot{N} \rangle^2\right].
\end{eqnarray}
The mean rate of the ticks is just twice the frequency of the oscillation, and its dispersion is dependent on the correlation function $\langle \dot{N}(0)\dot{N}(\tau)\rangle$. The latter describes the probability of having a tick at time $\tau$ given a tick at $t=0$. \par
Combining both distributions, the probability of finding the counter state $y$ at a given time $t$ can be identified as
\begin{eqnarray}
    P(y,t) = \sum_N P(y|N)P(N,t).
\end{eqnarray}
 The above distribution $P(y,t)$ is also a Gaussian with the following mean and variance:
\begin{equation}
    \begin{split}
        \langle y(t) \rangle &= \langle \dot{N} \rangle  \langle \Delta y \rangle t\\
        {\rm Var}[y(t)] &= 2\left[\sigma^2(\Delta y)\langle \dot{N} \rangle + D_N\langle \Delta y \rangle^2\right]t
        \label{eqn: counter_stat} 
    \end{split}
\end{equation}
  In the above regime of operation, the entropy production due to the biased CMOS inverter is given as  
\begin{eqnarray}
    \dot{\sigma}_{\rm{inv}} = k_B\langle \dot{N} \rangle  \langle \Delta y \rangle (\Delta V/2V_T).
\end{eqnarray}
Hence, the uncertainty product can be simplified as follows,
\begin{eqnarray}
    \mathcal{Q} = \left[\frac{\sigma^2(\Delta y)}{\langle \Delta y \rangle}+ \frac{D_N}{\langle \dot{N} \rangle}\langle \Delta y \rangle\right]\frac{\Delta V}{V_T}.
\end{eqnarray}
 \par

\subsection{2-state regime}
For the complete single-electron regime, we also required the elementary voltage at the output to be $v_e^{\rm{out}}/V_T\gg1$ to restrict the output voltage $v(t)$ to 2 states $v_{\pm}$. In such a regime, the equilibrium distributions corresponding to the coarse-grained states $(+,-)$ (Eq.~\eqref{eqn: Peq_+},~\eqref{eqn: Peq_-}) are peaked around the two states. The normalization of the distribution is effectively due to only two terms, i.e. $\mathcal{N}= \sum_{v_e^{\rm{out}}} P_{\rm{eq}}^+(v)\approx P_{\rm{eq}}^+(v_-) +P_{\rm{eq}}^+(v_+) $. Hence, the normalized equilibrium probabilities for the coarse-grained states can be computed as follows:
\begin{eqnarray}
    P_{\rm{eq}}^+(v_-)&=& \frac{e^{-V_{\rm{dd}}/2}}{[2\cosh{V_{\rm{dd}}/2}]}, \hspace{0.5cm}
    P_{\rm{eq}}^+(v_+) = \frac{e^{V_{\rm{dd}}/2}}{[2\cosh{V_{\rm{dd}}/2}]}\nonumber\\
    P_{\rm{eq}}^-(v_-)&=&\frac{e^{V_{\rm{dd}}/2}}{[2\cosh{V_{\rm{dd}}/2}]},\hspace{0.5cm}
    P_{\rm{eq}}^-(v_+)= \frac{e^{-V_{\rm{dd}}/2}}{[2\cosh{V_{\rm{dd}}/2}]}\nonumber
\end{eqnarray}
Using the above distributions, we can compute the mean increment of the counter given a tick, given as
\begin{flalign}
    \langle \Delta y \rangle = \tanh\left(V_{\rm{dd}}/2V_T\right)\hspace{0.1cm}\rm{and}\hspace{0.1cm}
    \sigma^2(\Delta y) = \frac{1}{2\cosh^2(V_{\rm{dd}}/2V_T)}.
    \label{eqn: increment_toy}
\end{flalign}
An important observation in these statistics is that the mean $\langle \Delta y \rangle$ increases to $1$ and the variance $\sigma^2(\Delta y)$ decreases with increasing voltage difference $\Delta V$. Therefore, the counting process becomes very precise by increasing the dissipation in the clock counter. The uncertainty product $\mathcal{Q}$ therefore simplifies to Eq.~\eqref{eqn: Q_toymodel}.

\subsection{Macroscopic counter state}
For the macroscopic limit in the output voltage of the biased CMOS inverter ($v_e^{\rm{out}}/V_T\to0$), the number of charges for a finite voltage will also be macroscopic. Even in this regime, the output voltage will relax to the same coarse-grained equilibrium distributions. But now the summation in the normalization can be approximated by an integral over the macroscopic number of states, i.e. $\mathcal{N}= \sum_{v_e^{\rm{out}}} P_{\rm{eq}}^+(v) \simeq \int_{-\infty}^{\infty} P_{\rm{eq}}^{\pm}(v) dv = \sqrt{2\pi v_e^{\rm{out}} V_T}.$ Hence, the corresponding equilibrium probabilities are 
\begin{eqnarray}
    P_{\rm{eq}}^{+} (v) &=& \frac{1}{\sqrt{2\pi v_e^{\rm{out}} V_T} }e^{-\frac{1}{(2v_e^{\rm{out}} V_T)}\left[v +V_{\rm{dd}}\right]^2}\label{eqn: Peq_+_macro}\\
    P_{\rm{eq}}^{-} (v) &=& \frac{1}{\sqrt{2\pi v_e^{\rm{out}}V_T}}e^{-\frac{1}{(2v_e^{\rm{out}} V_T)}\left[v - V_{\rm{dd}}\right]^2}.\label{eqn: Peq_-_macro}
\end{eqnarray}
The statistics of the increments given a tick are then given as,
\begin{flalign}
    \langle \Delta y \rangle = \left(2V_{\rm{dd}}/v_e^{\rm{out}}\right)\hspace{0.2cm}\text{and}\hspace{0.2cm}
    \sigma^2(\Delta y) =  2 V_T/v_e^{\rm{out}}.   
    \label{eqn: increment_macro}
\end{flalign}
For strong biasing $V_{\rm{dd}}/V_T>v_e^{\rm{out}}/V_T$, it takes $2V_{\rm{dd}}/v_e^{\rm{out}}$ transitions to change $v=\pm V_{\rm{dd}}\to \mp V_{\rm{dd}}$. The variance in increments is
controlled by the width of the Gaussian distributions, which depends only on the thermal voltage $V_T$.  Combining all of the above statistics, we can again obtain an analytical expression for the uncertainty product of the clock,
\begin{eqnarray}
        \mathcal{Q} = 2\left[1+ \left(\frac{D_N}{\langle \dot{N} \rangle}\right)\frac{2V_{\rm{dd}}^2}{v_e^{\rm{out}}V_T}\right].
\end{eqnarray}
Although the precision of the counter $\langle y\rangle^2/\sigma^2(y)$ increases with increasing applied voltage difference $\Delta V$, the variance $\sigma^2(y)$ is still independent of $\Delta V$. This is in contrast to the toy model, where both the precision and the variance of the increments decrease with a higher voltage difference (Eq.~\eqref{eqn: increment_toy}). Therefore, there is an ineffective conversion of the higher dissipation into the precision of the counting process at the macroscopic limit.

\section{Concusions and Discussion}
In this article, we have presented a thermodynamically consistent analysis of an electronic circuit inspired by the escapement mechanism used in mechanical clocks, based on the toy model in \cite{pietzonka2022classical}. 
In this circuit, the equilibrium stochastic oscillations of an RLC circuit drive the input of a biased CMOS inverter, where the accumulated current through the transistors acts as a discrete counter for timekeeping. 
In the single-electron regimes of the circuit, we showed that the dynamics of the counter converges to the toy model, which violates the TUR. Practically, this can be achieved only at ultra-low temperatures ($T\sim4K$) and with nanoscale components. In this regime, the state space of the CMOS inverter is confined to only two states, and the escapement coupling depends only on the coarse-grained position of the input oscillations. We also showed that those two constraints play a crucial role in the violation of the TUR. In the other regimes of operation, where either the coupling cannot be coarse-grained or the state space is meso/macroscopic, the TUR is restored. Hence, using an electronic implementation of an escapement clock, we show that the violation of the original TUR in underdamped systems requires specific conditions that can only be achieved in the single-electron regime of operation. Our work also sheds new light on the design of electronic circuits for timekeeping, utilizing thermal noise to reduce the thermodynamic cost.\par

The single-electron regimes, where the circuit converges to the toy model, require these solid-state devices and the RLC to be operated at extremely low temperatures. In this regime, the I-V characteristics used might also need to account for other dominant effects, such as quantum tunnelling, freezing of the charge carriers, etc. \cite{beckers2018characterization,beckers2021cryogenic}, which might also play an important role. These effects can significantly impact the performance of the escapement clock. Obtaining the coarse-grained coupling in these circuits is also a practical limitation that requires careful consideration. We achieved this by scaling the input voltage signal through reducing capacitances or operating at lower temperatures, which incurs no additional thermodynamic cost.  Alternatively, the coarse-grained coupling can be implemented through active amplification using an op-amp or using zero-crossing detectors \cite{irmak2011design} connected between the RLC and the CMOS inverter. All of these modules are dissipative and will contribute to the thermodynamic cost of running the clock. In addition, it is important to note that by scaling down the capacitances, the frequency of the pendulum scales up, as $\omega = 1/\sqrt{LC_{\rm{in}}}$. In order to maintain the clock at a finite frequency, the inductance has to be scaled up, which is not possible at the nanoscale. 

Similar circuits that incorporate some feedback of the counter dynamics back to the stochastic oscillations of the pendulum can be considered as Brownian versions of macroscopic mechanical clocks \cite{gorelik1998shuttle}. Although the oscillations in the pendulum can become more precise with feedback, it also drives the pendulum out of equilibrium with non-zero dissipation \cite{wachtler2019stochastic}. Thus, it is necessary to further explore the TUR bound on realistic systems that incorporate feedback. Since our circuit violates the TUR only at the single-electron regime, it is still an open question whether the TUR bound can be violated for macroscopic underdamped systems \cite{pearson2021measuring,milburn2020thermodynamics}.
\section{Acknowledgments}
A.G. thanks Patrick Pietzonka and Massimo Bilancioni for the discussions and valuable comments. This research was supported by the project INTER/FNRS/20/15074473 funded by F.R.S.-FNRS (Belgium) and FNR (Luxembourg).

\bibliography{references}

\begin{thebibliography}{53}
\bibitem [{2pr()}]{2problem}%
  \BibitemOpen
  \href@noop {} {\bibinfo {title} {{Consider} a sequence of voltages $\{
  v_{i-1},v_{i},v_{i+1},,v_{i+2}\}$ where $v_i$ is the voltages before $i^{th}$
  tick. {In} such a situation, the three consecutive increments are given as,
  ${\Delta y_{i-1} = s(i-1)[v_{i} - v_{i-1}}]$, ${\Delta y_i = s(i-1)[v_i -
  v_{i+1}]}$ and ${\Delta y_{i+1} = s(i-1)[v_{i+2} - v_{i+1}]}$, where we have
  used the alternating nature of the coarse-grained states
  $s(i-1)=s(i+1)=-s(i)$ and set $v_e=1$ for sake of simplilicity. {Since one of
  the voltage repeats for any two consecutive increments,} there will be a
  non-zero covariance between these terms. {These} covariances are given as
  ${\rm{{cov}}[\Delta y_i \Delta y_{i-1}] = \sigma^2(v_i)= \sigma_\pm^2(v)}$
  and $\rm{{Cov}}{[\Delta y_i \Delta y_{i+1}] = \sigma^2(v_{i+1})
  =\sigma_\mp^2(v)}$, where the sign will depend on the coarse-grained state
  $s(i)$. {Taking} the sum of all such covariances, the total variance of the
  counter given ${N\gg1}$ ticks is given as, $\sigma^2(y|{N}) =
  \sum_{i=0}^{{N}-1} \sigma^2({\Delta} y_i) +\sum_{i=0}^{{N}-1} \sum_{j=0,
  j\neq i}^{{N}-1} \rm{{Cov}}[{\Delta} y_i,{\Delta} y_j] \approx {N}
  \sigma^2({\Delta} y)+ {N}[\sigma^2_-({\Delta} y)+\sigma^2_+({\Delta} y)] =
  2{N} \sigma^2({\Delta} y)$.}}\BibitemShut {Stop}%
\end{thebibliography}

\appendix
\onecolumngrid
\counterwithin{figure}{section}

\section{Deterministic dynamics}
\label{apsec: deterministic}
The dynamics of the RLC can be independently determined from the dynamics in the CMOS inverter, as there is no current flow through the gate terminals of the transistors. Applying Kirchoff's voltage law around the RLC loop, we can obtain the following input voltage $V_{\rm{in}}(t)$ dynamics, 
\begin{eqnarray}
LC_{\rm{in}} \;\Ddot{V}_{\rm{in}}(t) + RC_{\rm{in}} \;\dot{V}_{\rm{in}}(t) + V_{\rm{in}}(t) =  0,
\label{eqn: det_input}
\end{eqnarray}
where $C_{\rm{in}}= C_i + 2 C_g$ is the effective capacitance at the input node of the inverter and $C_g$ is the gate-bulk capacitance of the transistors (Fig.~\ref{fig:Circuit}(c)).  As there is no voltage difference across the RLC, the input voltage $V_{\rm{in}}(t)$ will eventually relax to $V_{\rm{in}}^*=0$ in the steady state, which is the fixed point of the above dynamics. \par

Similarly, one can write the deterministic evolution of the output voltage of the CMOS inverter by applying Kirchoff's current law at the output node, given as
\begin{eqnarray}
    C_{\rm{out}} \frac{d v}{dt}= I_{p}(v, V_{\text{in}}; \,V_{\text{dd}}) - I_{n}(v, V_{\text{in}}; \,V_{\text{dd}}),
    \label{eqn: det_output}
\end{eqnarray}
where $C_{\rm{out}}= C_{\rm{b}} + 2 C_o$ is the effective output capacitance at the inverter output node and $C_o$ is the drain-source capacitance of the transistors (Fig.~\ref{fig:Circuit}(c)). The associated charge in the output node of the inverter is linearly related to the voltage as $q=C_{\rm{out}}v-C_{\rm{b}}V_{\rm{b}}.$

Since $V_{\rm{in}}(t)$ evolves independently, the steady-state output voltage $v^*$ can be obtained from Eq.~\eqref{eqn: det_output}, given its steady-state value $V_{\rm{in}}^*=0$. The symmetry of the currents, $I_{n}(v, V_{\text{in}}; \,V_{\text{dd}})=I_{p}(-v,-V_{\text{in}};\,V_{\text{dd}})$, implies the steady-state output voltage $v^*=0$, for any $V_{\rm{dd}}$. Therefore, the steady state of the deterministic dynamics (Eqs.~\ref{eqn: det_input}\&\ref{eqn: det_output}) is just $(v^*, V_{\rm{in}}^*) = (0, 0)$. To have sustained oscillations in the steady state of digital clocks (RLC/ crystal oscillators),  part of the power injected into the inverter is also fed back to the damped oscillator to drive the oscillations. Here, the thermal noise in the resistors probes the natural frequency of the oscillator to produce stochastic oscillations whose coherence time depends on the damping factor.

\section{Coarse-graining the jump rates of the transistors}
\label{apsec: Coarse-grained-coupling}
In this section, we will argue that for $v_e^{\rm{in}}/V_T\gg1$, the dependence of the jump rates $\lambda_\rho(t)$ on the input voltage $V_{\rm{in}}(t)$ can be coarse-grained so as to depend only on its sign. More specifically, the counter dynamics $y(t)$ converges to an auxiliary dynamics with the following escapement coupling : $\lambda_p^{\pm}(t)\propto \Theta(-V_{\rm{in}}(t)/V_T)$ and $\lambda_n^{\pm}(t)\propto \Theta(V_{\rm{in}}(t)/V_T)$, where $\Theta(x)$ is a Heaviside function such that $\Theta(x) = 1$ for $x\ge0$ and $0$ elsewhere. With such coarse-grained rates, the transistors behave exactly like switches, i.e. for $V_{\rm{in}}(t)\ge0$, only the jumps in nMOS $(\pm n)$ are allowed, whereas the jumps in pMOS $(\pm p)$ are shut off, and vice versa for $V_{\rm{in}}(t) < 0$.

\begin{figure*}[ht!]
    \centering
    \includegraphics[trim=0 0 0 0, clip,scale=0.5]{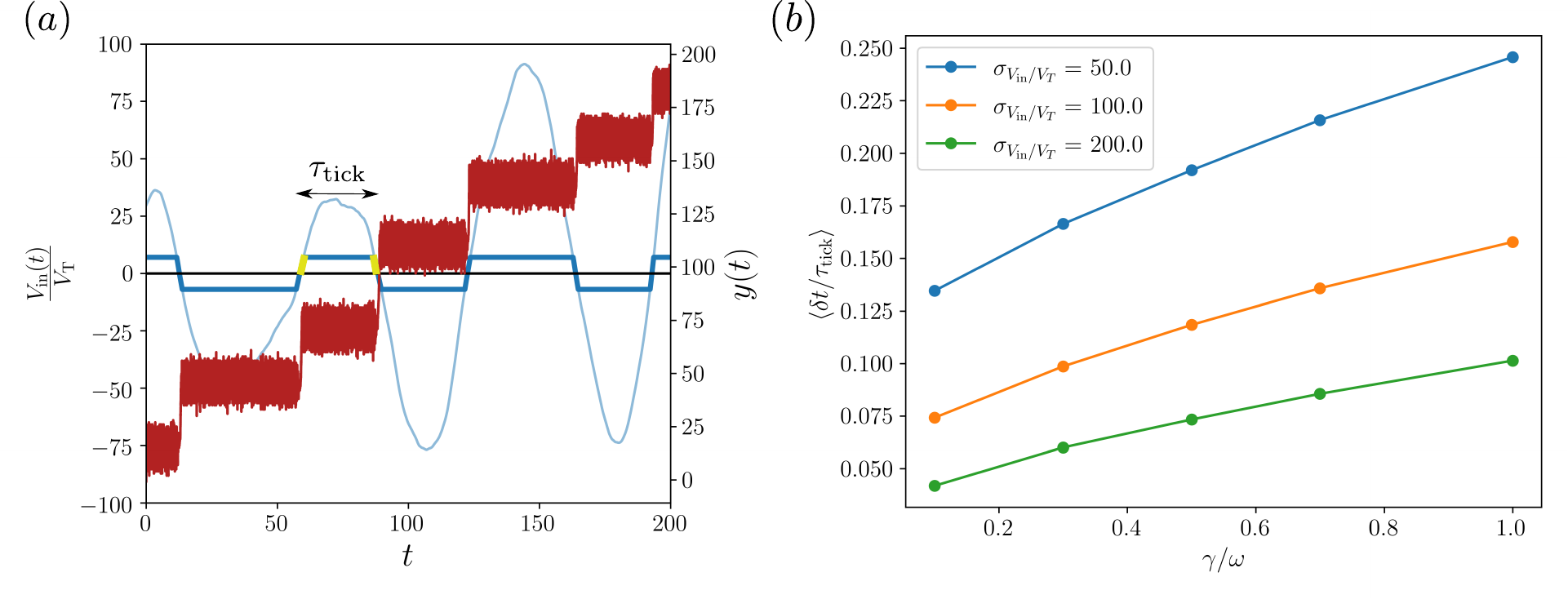}
    \caption{(a) A sample stochastic trajectory of the input voltage $V_{\text{in}}(t)$ (blue), the coarse-grained input voltage trajectory (dark blue) ($V_{\rm{in}}^{\rm{th}}/V_T=7$) and the trajectory of the counter-observable $y(t)= N_{p}(t) + N_{n}(t)$  (red). For a sample time interval between ticks $\tau_{\rm{tick}}$, the part of the trajectory spent in the threshold range $[- V_{\rm{in}}^{\rm{th}}, V^{\rm{th}}_{\rm{in}}]$ is highlighted in yellow, and $\delta t$ is the corresponding time duration. Parameters: $\omega \equiv 1/\sqrt{LC_{\rm{in}}} = 0.1 \tau_0^{-1}, \sigma_{V_\text{in}/V_T} = 50, v_e^{\rm{out}}/V_T = 0.1$ and $C_bV_b = -q_e/2.$ (b) The average $\delta t/\tau_{\rm{tick}}$ for different values of damping $\gamma/\omega$ and standard deviation of input voltage $\sigma_{V_\text{in}/V_T} = \sqrt{v_e^{\rm{in}}/V_T}$. Parameter: $\omega \equiv 1/\sqrt{LC_{\rm{in}}} = 0.1 \tau_0^{-1}.$}
    \label{fig: app_traj_threshold}
\end{figure*}

As explained in the main text, we first note that the jump rates for transistors, obtained from the I-V characteristics, have the following input voltage dependence: $\lambda_\pm^n \propto e^{V_\text{in}(t)/V_T}$ for the nMOS transistor and $\lambda_\pm^p \propto e^{-V_\text{in}(t)/V_T}$ for the pMOS transistor. Therefore, the transistors will behave effectively like switches for $|V_\text{in}(t)/V_T| \gg 1$ with only one of the transistors active. For $v_e^{\rm{in}}/V_T\gg1$, this condition is achieved during most of the period of an oscillation, except for the time spent around the zero crossing. For a quantitative analysis, we can consider the time $\delta t$ spent by the input signal $V_{\rm{in}}(t)$ in a range $[-V_{\rm{in}}^{\rm{th}}, V^{\rm{th}}_{\rm{in}}]$, where $V^{\rm{th}}_{\rm{in}}$ is a threshold voltage which is in principle arbitrary. 
This threshold voltage can be chosen so that, when $|V_\text{in}(t)| > V_\text{in}^\text{th}$, the timescale for jumps in the inactive transistor is much larger than the time between ticks, i.e. $e^{V_{\rm{in}}^{\rm{th}}/V_T}\tau_0 \gg \langle \tau_{\rm{tick}} \rangle$.
Both $\delta \tau$ and $\tau_\text{tick}$
are random quantities. As shown in Fig.~\ref{fig: app_traj_threshold}. (b), the average of the fraction $\delta \tau/\tau_{\rm{tick}}$ can be made arbitrarily small by considering larger values of $v_e^{\rm{in}}/V_T$ (as $\sigma_{V_\text{in}/V_T} =\sqrt{v_e^{\rm{in}}/V_T}$), for any damping $\gamma/\omega$. In the case of low damping $\gamma \ll \omega$, one can show that $\langle \delta \tau/\tau_{\rm{tick}} \rangle \propto (V_{\rm{in}}^{\rm{th}}/V_T)/\sqrt{v_e^{\rm{in}}/V_T}$, capturing the inverse relationship to $v_e^{\rm{in}}$. This approximate relationship is obtained by performing a linear interpolation assuming that the amplitude of the input voltage $|V_{\rm{in}}^{\rm{amp}}/V_T| \sim \sigma_{V_{\rm{in}}/V_T}$ is reached at $t\sim \langle\tau_{\rm{tick}}\rangle/2 = \pi/(2\omega)$, given that $V_{\rm{in}}/V_T=0$ at $t=0$ (the zero crossing). Thus, if $\delta\tau \ll \tau_0 \ll \langle \tau_{\rm{tick}} \rangle$, where $\tau_0$ is the natural timescale of the transistor defined in the main text, the transistors will effectively never see the input voltage around 0. \par
Now we focus on the independence of the auxiliary dynamics on the magnitude of the input voltage $|V_{\rm{in}}(t)/V_T|$. Under the above condition, the output voltage $v(t)$ quickly equilibrates with the corresponding voltage source (at voltage $V_{\rm{dd}}$ or $-V_{\rm{dd}}$) of the conducting transistor, once zero-crossing occurs. Thus, after that quick equilibration, the electric current has a zero mean during most of the period between ticks. This implies that the counter dynamics within a period does not contribute to the progress of the counter state $y(t)$ irrespective of the magnitude of $V_{\rm{in}}(t)$ (Fig.~\ref{fig: app_traj_threshold}. (a)). Therefore, there are advances in the counter observable $y(t)$ only at the moments of zero crossing or tick events due to the relaxation of the output voltage to the new equilibrium associated with the newly activated transistor.

\section{Precision of the number of zero-crossing in the RLC}
\label{apsec: precision_ticks}
In this section, we obtain the expression for the precision of the number of zero crossings of the input voltage in the RLC. Since the stochastic dynamics of the RLC is that of an underdamped harmonic oscillator, we can cast Eq.~\eqref{eqn: Langevin_input} into an adimensional form in the phase space of $(x\equiv \sqrt{C_{\rm{in}}/k_B T}\, V_{\rm{in}}, v\equiv \sqrt{L C_{\rm{in}}^2/k_B T}\, \dot{V}_{\rm{in}})$. After rescaling time to dimensionless time as $t \to t/\sqrt{L C_{\rm{in}}}$, we get the following equations of motion
\begin{eqnarray}
    \partial_t x &=&v\\
    \partial_t v &=& -x - \Tilde{\gamma}v + \Tilde{\xi}(t),
\end{eqnarray}
where $\Tilde{\xi}(t)$ is Gaussian white noise with correlations, $\Tilde{\xi}(t)\Tilde{\xi}(t')= 2\Tilde{\gamma}\delta(t-t')$ and effective damming coefficient $\Tilde{\gamma}=\gamma/m\omega$. It should be noted that the above dynamics is only dependent on this damping factor $\Tilde{\gamma}$, while all other parameters scale the trajectory. The equilibrium distribution of the above dynamics correspond to,
\begin{eqnarray}
    P_{\rm{eq}}(x,v) = \frac{1}{2\pi}e^{-(x^2+v^2)/2}.
\end{eqnarray}
Note that $(x^2+v^2)/2 \equiv (L C_{\rm{in}}^2 \dot{V}_{\rm{in}}^2+ C_{\rm{in}} V_{\rm{in}}^2)/(2k_B T) $ is the rescaled energy stored in the RLC, thus obtaining the correct Gibbs state.\par
We are interested in computing the dispersion of the observable $N(t)=\int_0^t|v|\delta(x)$, which counts the number of zero crossings (ticks) up to time $t$. Below, we sketch the derivation done in \cite{pietzonka2022classical} and also correct for a typo in the final expression. For long times $t\gg1$, the distribution of $N(t)$ can be assumed to be Gaussian with mean $\langle \dot{N} \rangle t$ and dispersion $2 D_N t$ using the central limit theorem. 
We can compute both at equilibrium using the result from \cite{seifert2010generalized} as
\begin{eqnarray}
    \langle \dot{N} \rangle &=& \int dx dv\; P_{\rm{eq}}(x,v) |v| \delta(x) = 1/\pi\\
    D_N &=& \frac{\langle \dot{N} \rangle}{2}+\int_{0^+}^{\infty}d\tau\left[\langle \dot{N}(0)\dot{N}(\tau)\rangle - \langle \dot{N} \rangle^2\right].
\end{eqnarray}
The dispersion $D_N \equiv \lim_{t\to\infty}1/(2t)\int_0^t dt_1 \int_0^t dt_2 \langle [\dot{N}(t_1)- \langle \dot{N} \rangle] [\dot{N}(t_2)-\langle \dot{N} \rangle ]\rangle$ captures the correlations in the underdamped dynamics of the RLC, and the first term is the result of self-correlations in the ticks. Since the underdamped harmonic oscillator is a linear system, the propagator $P(x,v,t|x_0,v_0,0)$ is also Gaussian. Using the Gaussian propagator, the correlation function $\langle \dot{N}(0)\dot{N}(\tau)\rangle$ can be exactly computed, as shown in the supplementary material of \cite{pietzonka2022classical}, giving us
\begin{eqnarray}
    \langle \dot{N}(0)\dot{N}(\tau)\rangle = \frac{1}{\pi^2\sqrt{\det \bf{\sigma}}\det \bf{\sigma}}\left[1+\frac{|\sigma_{12}|}{\sqrt{\det \bf{\sigma}}}\arctan\frac{|\sigma_{12}|}{\sqrt{\det \bf{\sigma}}}\right].\nonumber
    \end{eqnarray}
In the above expression, the matrix $\bf{\sigma}$ is given as
\begin{eqnarray}
    \bf{\sigma}\equiv \begin{pmatrix} 1 & 0\\ 0 & 0    \end{pmatrix} + \bf{B}^T\bf{\sigma}^{-1}\bf{B}
\end{eqnarray}
with 
\begin{eqnarray}
    \bf{B} &\equiv& \begin{pmatrix} \frac{e^{-\Tilde{\gamma}t/2}}{\Tilde{\omega}}\sin(\Tilde{\omega}t) & 0\\  \frac{e^{-\Tilde{\gamma}t/2}}{\Tilde{\omega}}[\Tilde{\omega}\cos(\Tilde{\omega}t)-(\Tilde{\gamma}/2)\sin(\Tilde{\omega}t)] & 1   \end{pmatrix},\nonumber\\
    \bf{\sigma}(t) &\equiv& \begin{pmatrix} 1 & 0\\ 0 & 1  \end{pmatrix} + \frac{e^{-\Tilde{\gamma}t/2}}{2\Tilde{\omega}^2}\left[\begin{pmatrix} -2 & \Tilde{\gamma}\\ \Tilde{\gamma} & -2  \end{pmatrix} + \begin{pmatrix}  \Tilde{\gamma}/2 & -1\\ -1 &  \Tilde{\gamma}/2  \end{pmatrix} \Tilde{\gamma}\cos(2 \Tilde{\omega}t) + \begin{pmatrix} -1 & 0\\ 0 & 0    \end{pmatrix}\Tilde{\gamma}\Tilde{\omega}\sin(2 \Tilde{\omega}t)\right]
\end{eqnarray}
where $\Tilde{\omega}=\sqrt{1-\Tilde{\gamma}^2/4}$ is the shifted frequency of the oscillator due to the damping. Since the above expressions are lengthy and cannot be further simplified, the integral in $D_N$ is computed numerically to obtain the precision $D_N/\langle \dot{N} \rangle$ of the ticks.

\section{Auxiliary dynamics for continuous coupling}
\label{apsec: Approximate counter dynmaics}
The operation of circuit with the RLC having macroscopic capacitance implies that the thermal oscillations in the RLC have small amplitudes, since $v_e^{\rm{in}}/V_T<1$. Therefore, the dynamics of the counter $y(t)$ will now explicitly depend on the values of the input voltage $V_{\rm{in}}(t)$. To capture the qualitative behaviour in this regime, we consider an approximate counter dynamics $y(t)$ assuming a time-scale separation between the RLC and the CMOS inverter. Here, we construct a Langevin dynamics for the counter with a drift $\mu(V_{\text{in}}(t))$ and a diffusion $\Tilde{\sigma}(V_{\rm{in}}(t))$ coefficient. Coupled with the underdamped dynamics for the input voltage $V_{\rm{in}}(t)$, we get an effective Langevin dynamics for the counter $y(t)$ in an extended space $(y, V_{\rm{in}}(t))$, given as
\begin{flalign}
LC_{\rm{in}}& \Ddot{V}_{\rm{in}}(t) + RC_{\rm{in}} \;\dot{V}_{\rm{in}}(t) + V_{\rm{in}}(t) =   \xi(t), \nonumber\\
\dot{y}(t) &= \mu(V_{\text{in}}(t))  + \Tilde{\sigma}(V_{\rm{in}}(t)) \,\eta_t,
\label{apeqn: Gaussian_timescale}
\end{flalign}
where $\eta(t)$ is a Gaussian white noise with zero mean and unit variance. The drift and diffusion coefficients are assumed to be equal to the long time limit of the counter statistics in the CMOS inverter for a \textit{fixed} $ V_{\rm{in}}$. Specifically, the drift coefficient can then be computed as $\mu(V_{\text{in}})=\lim_{\tau \to \infty}(\langle y(\tau) \rangle_{V_{\rm{in}}}/\tau)$ and the diffusion coefficient is computed as $\Tilde{\sigma}(V_{\rm{in}}) = \lim_{\tau \to \infty}\sqrt{\text{Var}[y(\tau)]_{V_{\rm{in}}}/\tau}$. This approximate dynamics provides a faithful and numerically inexpensive alternative to the time-dependent Gillespie simulations of the circuit when there is timescale separation between the RLC and the CMOS inverter. It is also important to note that the above dynamics do not capture the regime with coarse-grained rates, as it still assumes time-scale separation near $V_{\rm{in}}=0$, which is no longer true in that case (See Appendix.~\ref{apsec: Coarse-grained-coupling}). Below, we will compute the drift and diffusion coefficient in the 2-state and macroscopic state space regimes. \par

\subsection{Two-state limit}
As shown in the main text, if we consider the limit $v_e^{\rm{out}}/V_T\gg1$, the state space of the output voltage/charge can be restricted. By tuning the biasing circuit such that $C_b V_b = -(n-1/2)q_e$, it will be restricted to two states $v_- \equiv (n-1) q_e$ and $v_+ \equiv n q_e $. The Poisson transition rates between these states due to the transistors in the inverter (assuming the slope factor $n=1$) are given as,
\begin{eqnarray}
\lambda_{p}^+(v_-\to v_+) &=& (1/\tau_0)e^{(V_\text{dd} -
V_\text{in})/V_T} ,\qquad\qquad \;\lambda_{n}^+(v_+\to v_-) = (1/\tau_0)e^{(V_\text{dd} +
V_\text{in})/V_T}, \\
\lambda_{p}^-(v_+\to v_-) &=& (1/\tau_0)e^{-
V_\text{in}/V_T}; \qquad\qquad \qquad  \lambda_{n}^-(v_-\to v_+) = (1/\tau_0)e^{
V_\text{in}/V_T} ;
\end{eqnarray}
Using the methods of full counting statistics \cite{bagrets2003full,esposito2009nonequilibrium}, one can exactly compute all the cumulants of the counter observable $y(t)$ in this two-state system. The tilted generator corresponding to the counter observable $y(t) \equiv  N_{p}(t) + N_{n}(t)$ for any fixed input voltage $V_\text{in}/V_T$ is given as,
\begin{eqnarray}
\hat{L}_{\xi} = \begin{bmatrix}
-(\lambda_{p}^+ + \lambda_{n}^-) & (\lambda_{p}^-e^{-\xi}  + \lambda_{n}^+e^{\xi}) \\
(\lambda_{p}^+e^{\xi}  + \lambda_{n}^-e^{-\xi})  & -(\lambda_{p}^- + \lambda_{n}^+)
\end{bmatrix}
\end{eqnarray}
The scaled cumulant generating function (SCGF) $S(\xi)=\lim_{t\to \infty}(1/t)\log\langle e^{ \xi \, y(t)}\rangle$ is the eigenvalue with the largest absolute value of the titled generator $\hat{L}_{\xi}$. Similar calculations for the current statistics of a single current in a two-level system can be found in the Appendix.~F of \cite{gopal2022large}. From the SCGF, the drift $\mu(V_{\text{in}})$ and the diffusion $\Tilde{\sigma}(V_{\rm{in}})$ coefficient for the 2-state limit can be obtained from the following expressions,
\begin{eqnarray}
\mu(V_{\text{in}}) &=& \frac{\partial S(\xi)}{\partial\xi}\bigg|_{\xi=0} = \frac{1}{\tau_0}\frac{2e^{ V_\text{in}/V_T} (1-e^{-V_\text{dd}/V_T})}{(1+e^{ 2V_\text{in}/V_T})} \label{apeqn : 2state_av}\\
\Tilde{\sigma}(V_{\rm{in}}) &=& \sqrt{\frac{\partial^2 S_p(\xi)}{\partial\xi^2}}\bigg|_{\xi=0} = \frac{1}{\sqrt{\tau_0}}\left[\frac{8e^{-2V_\text{dd}/V_T+V_\text{in}/V_T}(2e^{2V_\text{in}/V_T}+(1+e^{4V_\text{in}/V_T})\cosh{(V_\text{dd}/V_T)})}{(1+e^{V_\text{dd}/V_T})(1+e^{2V_\text{in}/V_T})^3}\right]^{1/2}
\label{apeqn : 2state
_var}
\end{eqnarray}
\subsection{Macroscopic limit}
For the macroscopic limit in the output state space, we will be using the framework used in \cite{gopal2022large} to obtain the drift and diffusion coefficients. Taking the macroscopic limit $v_e^{\rm{out}}/V_T \to 0$ also corresponds to the low-noise limit in these electronic circuits, as the fluctuations scale as $v_e^{\rm{out}}/V_T$. In this limit, the probability of individual trajectories for such Markov jump processes can be obtained using the Martin-Siggia-Rose (MSR) path integral construction \cite{martin1973statistical,lazarescu2019large,tomassothesis}, and identifying the dominant trajectory. The generating function for counter statistics $Z(\xi,t)=\langle e^{ \xi y(t)}\rangle$ in this representation to the dominant order in $v_e^{-1}$, has the following form:
\begin{equation}
    Z(\xi, t) \!=\! \! \int\!\! \mathcal{D}v \mathcal{D}p \:\: e^{(1/v_e^{\rm{out}}) \! \!\int_0^t d\tau \left[ -p(\tau) \dot v(\tau) + 
    H_\xi(v(\tau), p(\tau))\right]} P_0(v(0)),
    \label{eq:current_path_int}
\end{equation}
where  $p(t)$ is the auxiliary field. This field plays the role of a conjugated momentum and is obtained when the Fourier transform of Poisson noise is taken. $P_0(v(0))$ is an initial probability distribution, and $H_\xi(v,p)$ is the \emph{biased} Hamiltonian
\begin{eqnarray}
    H_\xi(v,p)=\sum_{\rho}\left(e^{\Delta_{\rho} p +s(\rho)\xi}-1\right)\omega_{\rho}(v).
    \label{eqn: biased_H}
\end{eqnarray}
In the above biased Hamiltonian, $\omega_\rho(v) \equiv \lim_{v_e\to0} v_e\lambda_\rho(v,v_e)$ are the rescaled Poisson rates and $s(\rho)$ is the sign function which outputs $\pm 1$ corresponding to the $\pm$ jumps in the devices. Taking the long-time limit, the SCGF $S(\xi)$ can then be computed as
\begin{eqnarray}
     S(\xi)=\frac{1}{v_e^{\rm{out}}}\max_{v^*_\xi,p^*_\xi}\left\{ H_\xi(v^*_\xi,p^*_\xi)\right\},
     \label{eqn: SCGF_Hamiltonian}
\end{eqnarray}
where the maximization is done over the fixed points $\{v^*_\xi,p^*_\xi\}$ of the Hamiltonian dynamics $\dot{v_\xi}=\partial_p H_\xi(v,p)$ and $\dot{p_\xi}=-\partial_v H_\xi(v,p)$.
As the output voltage $v$ is related to the output charge $q$, by $v = (q + C_{\rm{b}}V_{\rm{b}})/=C_{\rm{out}}$, the rates in the voltage space are independent of the biasing (See Eq.~\ref{eqn: rates_p} and ~\ref{eqn: rates_n}) and the rescaled rates $\omega_\rho(v)$ are also the same as with the unbiased CMOS inverter. Therefore, the drift $\mu(V_{\text{in}}) = (\partial S(\xi)/\partial\xi)|_{\xi=0}$ and the diffusion $\Tilde{\sigma}(V_{\rm{in}}) = \sqrt{\partial^2 S_p(\xi)/\partial\xi^2}|_{\xi=0}$ coefficient can be semi-analytically computed using the results obtained for the unbiased inverter in \cite{gopal2022large}. 


\section{Stochastic simulation method}
Since the dynamics of the RLC occurs independently of the transitions in the CMOS inverter, we can hence independently solve the underdamped dynamics in the RLC numerically using traditional Runge-Kutta (RK) approaches \cite{greiner1988numerical}. Here, we obtained the stochastic trajectories of the RLC using Heun's method, which corresponds to a second-order RK approach. These trajectories provide the explicit time dependence for the Poisson rates in the CMOS inverter (Eq.~\eqref{eqn: rates_n}). Now, the stochastic trajectories of changes in the output voltage $v(t)$ of the biased CMOS inverter are numerically obtained using the time-dependent variant of the Gillespie algorithm \cite{gillespie1978monte,anderson2007modified}. The algorithm is implemented for our circuit as following:
\begin{itemize}
    \item Initialize the output voltage at some initial voltage $v(t_c=0)=v_0$
    \item To compute the next jump time, solve the following non-linear integral equation,
    \begin{eqnarray}
        \int_{t_c}^{t_c+\Delta}d\tau\sum_{\rho}\lambda_\rho(v(t_c),\tau) = \ln(1/r_1),
    \end{eqnarray}
    where $r_1\sim\rm{Uniform}[0,1]$ is a uniform random variable.
    \item Generate another uniform random number $r_2\sim\rm{Uniform}[0,1]$.
    \item Choose the Poisson process $\rho'$ associated with that jump, which satisfies the following condition:
    \begin{eqnarray}
        \sum_{\rho=1}^{\rho'-1}\lambda_\rho(v(t_c),t_c+\Delta)\le r_2\sum_{\rho=1}^{M}\lambda_\rho(v(t_c),t_c+\Delta)\le \sum_{\rho=\rho'}^{M}\lambda_\rho(v(t_c),t_c+\Delta), 
    \end{eqnarray}
where the different Poisson processes $\rho$ are given a dictionary ordering from 1 to $M$. In the case of the above circuit, the total number of processes is $M=4$.
\item Update the output voltage $v(t_c+\Delta)=v(t_c)+\Delta_{\rho'} v_e^{\rm{out}}$ according to the process $\rho'$, and the current time $t_c=t_c+\Delta$.
\item Repeat from step 2.
\end{itemize}
The advantage of the above Gillespie algorithm compared to the first-reaction method \cite{anderson2007modified} is that it requires only the generation of two random numbers for every jump. This method also is faster than the modified next-reaction methods, which require only one random number per jump for our nonlinear rates. The most numerically expensive step in the above algorithm with nonlinear rates is finding $\Delta$, which needs to be computed for each reaction in the modified next-reaction approach.

\section{Counter Uncertainty and Entropy production for different regimes}
\label{ap: spliting}
As discussed in the main text, the focus of this article is to quantify the uncertainty product $\mathcal{Q}$ and its connection with the thermodynamic uncertainty relation (TUR) for the different regimes of operation of the clock circuit. The uncertainty product $\mathcal{Q}$ is defined as the long time limit of the product of the uncertainty of the counter observable $\text{Var}[y(t)]/\langle y(t) \rangle^2$ and the total entropy production $\Dot{\sigma}t/k_b$,  
\begin{eqnarray}
    \mathcal{Q}\equiv\lim_{t\to\infty}\frac{{\rm Var}[y(t)]}{\langle y(t) \rangle^2} \: \frac{\dot{\sigma} t}{k_b}.
 \end{eqnarray}
In the long time limit, the counter-observable statistics are such that asymptotically its mean and variance grow linearly with time, i.e. $\langle y(t) \rangle \propto t$ and $\text{Var}[y(t)] \propto t$ \cite{gopal2022large}.  We define the scaled uncertainty for the counter observable as $\text{Var}[y(t)]t/\langle y(t) \rangle^2$, which is independent of time. Hence, the uncertainty product $\mathcal{Q}$ can be expressed as the product of the scaled uncertainty and the entropy production rate $\Dot{\sigma}/k_b$. This splitting allows us to understand the time-independent dependence of the uncertainty and the thermodynamic cost on the system parameters ($\Delta V, \gamma/m, v_e^{\text{in/out}} $). \par
\begin{figure}[h!]
    \centering
    \hspace{-1.3cm}
    \subfloat[]{{\includegraphics[trim=0 0 425 0, clip,scale=0.5]{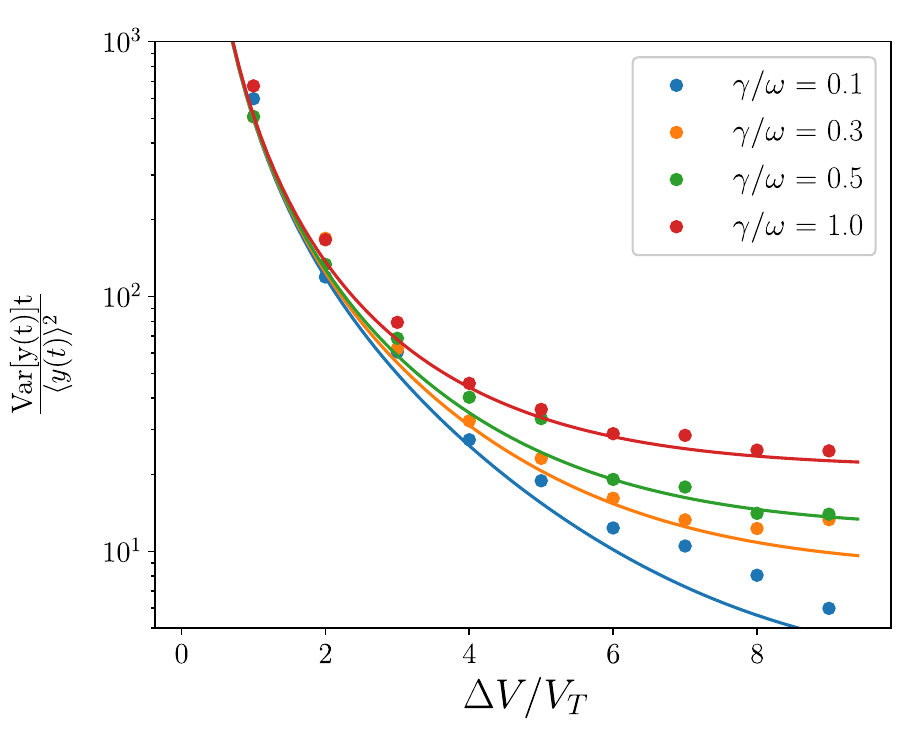} }}%
    \subfloat[]{{\includegraphics[trim=440 0 0 0, clip,scale=0.5]{Figure_app_2.pdf} }}%
    \caption{(a) The uncertainty $\frac{\text{Var}[y(t)]t}{\langle y(t)\rangle^2}$  and (b) the entropy production rate $\Dot{\sigma}$ of the counter $y(t)$ in the 2 state regime with coarse-grained coupling.
    It is plotted as a function of the voltage difference $\Delta V$ for different damping rates of the RLC circuit. Markers are obtained from Gillespie circuit simulations and solid curves are obtained by applying Eq.~\eqref{eqn: increment_toy} in Eq.~\eqref{eqn: counter_stat}. Parameters: $\omega \equiv 1/\sqrt{LC_{\rm{in}}} = 0.1\tau_0^{-1}, V_{\rm{in}}^{\rm{th}}/V_T=10, v_e^{\rm{out}}/V_T = 10$ and $C_bV_b = -q_e/2$.}
    \label{fig: app_2state_coarse}
\end{figure}
\begin{figure}[h!]
    \centering
    \hspace{-1.3cm}
    \subfloat[]{{\includegraphics[trim=0 0 425 0, clip,scale=0.5]{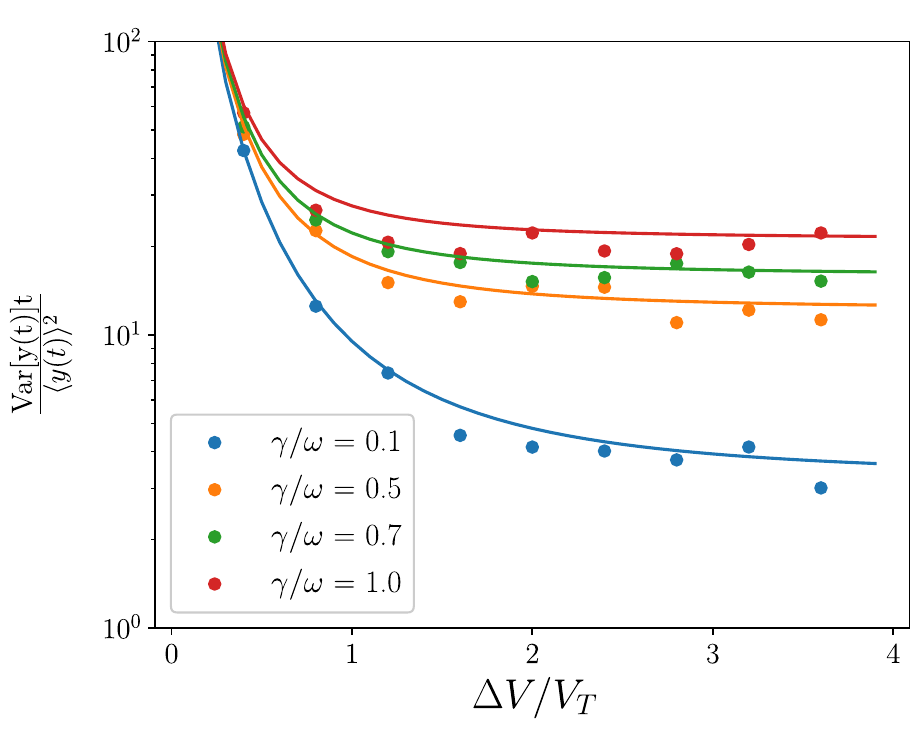} }}%
    \subfloat[]{{\includegraphics[trim=440 0 0 0, clip,scale=0.5]{Figure_app_3.pdf} }}%
    \caption{(a) The uncertainty $\frac{\text{Var}[y(t)]t}{\langle y(t)\rangle^2}$  and (b) the entropy production rate $\Dot{\sigma}$ of the counter $y(t)$ in the macroscopic state space with coarse-grained coupling.
    It is plotted as a function of the voltage difference $\Delta V$ for different damping rates of the RLC circuit. The markers are obtained from the Gillespie simulations of the circuit and the solid curves are obtained by applying Eq.~\eqref{eqn: increment_macro} in Eq.~\eqref{eqn: counter_stat}. Parameters: $\omega \equiv 1/\sqrt{LC_{\rm{in}}} = 0.1\tau_0^{-1}, V_{\rm{in}}^{\rm{th}}/V_T=10, v_e^{\rm{out}}/V_T = 0.05$ and $C_bV_b = -q_e/2$.}
    \label{fig: app_macroscopic_coarse}
\end{figure}
\begin{figure}[h!]
    \centering
    \hspace{-1.3cm}
    \subfloat[]{{\includegraphics[trim=0 0 350 0, clip,scale=0.6]{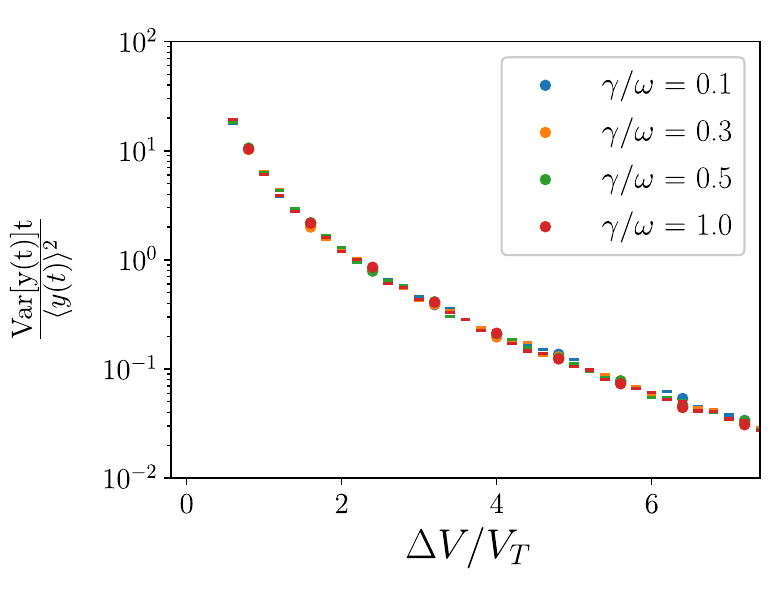} }}%
    \subfloat[]{{\includegraphics[trim=370 0 0 0, clip,scale=0.6]{Figure_app_4.pdf} }}%
    \caption{(a) The uncertainty $\frac{\text{Var}[y(t)]t}{\langle y(t)\rangle^2}$  and (b) the entropy production rate $\Dot{\sigma}$ of the counter $y(t)$ in the 2 state limit and regime where the rates cannot be coarse-grained.
    It is plotted as a function of the voltage difference $\Delta V$ for different damping rates of the RLC circuit. The markers are obtained from the Gillespie simulations with the time-dependent rates of the circuit. The dashed curves are computed using the approximate dynamics of Eq.~\eqref{apeqn: Gaussian_timescale}.  Parameters: $\omega \equiv 1/\sqrt{LC_{\rm{in}}} = 0.1$, $v_e^{\rm{in}}/V_T = 0.1$, $v_e^{\rm{out}}/V_T = 10$ and $C_bV_b = -q_e/2$.}
    \label{fig: app_2state_true}
\end{figure}
\begin{figure}[h!]
    \centering
    \hspace{-1.3cm}
    \subfloat[]{{\includegraphics[trim=0 0 420 0, clip,scale=0.5]{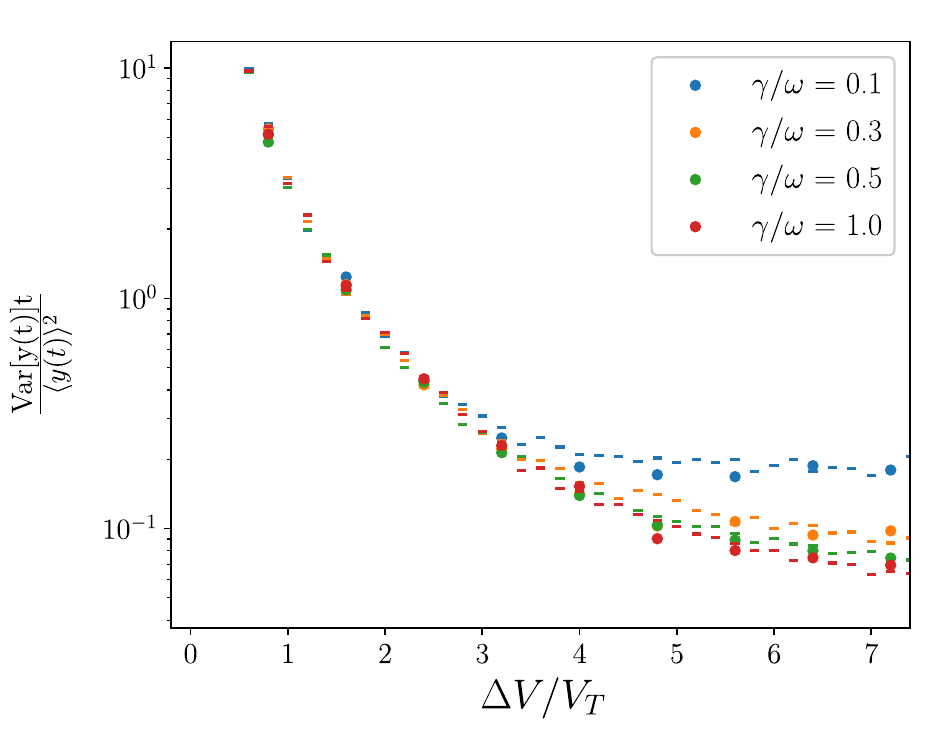} }}%
    \subfloat[]{{\includegraphics[trim=450 0 0 0, clip,scale=0.5]{Figure_app_5.pdf} }}%
    \caption{(a) The uncertainty $\frac{\text{Var}[y(t)]t}{\langle y(t)\rangle^2}$  and (b) the entropy production rate $\Dot{\sigma}$ of the counter $y(t)$ in the macroscopic operating regimes, i.e. macroscopic state space and jump rates which cannot be coarse-grained.
    It is plotted as a function of the voltage difference $\Delta V$ for different damping rates of the RLC circuit. The markers are obtained from the Gillespie simulations with the time-dependent rates of the circuit. The dashed curves are computed using the approximate dynamics of Eq.~\eqref{apeqn: Gaussian_timescale}.  Parameters: $\omega \equiv 1/\sqrt{LC_{\rm{in}}} = 0.1$, $v_e^{\rm{in}}/V_T = 0.1$, $v_e^{\rm{out}}/V_T = 0.1$ and $C_bV_b = -q_e/2$.}
    \label{fig: app_macroscopic_true}
\end{figure}

Below we plot the scaled uncertainty $\text{Var}[y(t)]t/\langle y(t) \rangle^2$ and the entropy production rate $\Dot{\sigma}/k_b$ separately, as a function of applied voltage $\Delta V$ for different damping factors $\gamma/m$. We do the same for the different regimes of operation of the circuit, as discussed in the main text: (1). 2-state limit ($v_e^\text{out}/V_T \gg 1$) and coarse-grained rates ($v_e^\text{in}/V_T \gg 1$) (Fig.~\ref{fig: app_2state_coarse}), (2). macroscopic state space ($v_e^\text{out}/V_T < 1$) with coarse-grained rates (Fig.~\ref{fig: app_macroscopic_coarse}), (3). 2-state limit but without coarse-grained rates ($v_e^\text{in}/V_T < 1$) (Fig.~\ref{fig: app_2state_true}), and finally (4). macroscopic regime of the entire circuit ($v_e^\text{in}/V_T < 1$ and $v_e^\text{out}/V_T < 1$) (Fig.~\ref{fig: app_macroscopic_true}).\par
As seen in the figures, the entropy production rate $\Dot{\sigma}/k_b = \langle y(t)/t \rangle (\Delta V/V_T)$ is independent of the damping factor $\gamma/\omega$. At equilibrium, the mean counter observable $\langle y(t) \rangle$ is computed using the Gibbs probability density of the oscillator, which is independent of the damping $\gamma$. But, as seen with coarse-grained rates, the scaled variance $\text{Var}[y(t)]t/\langle y(t) \rangle^2$ is proportional to dispersion in the ticks (Appendix.~\ref{apsec: precision_ticks}), which decreases with decreasing damping factor $\gamma/\omega$. But when the rates cannot be coarse-grained, the scaled uncertainty has the opposite behaviour (increases with decreasing damping factor $\gamma/\omega$) (See Fig.~\ref{fig: app_2state_true} and Fig.~\ref{fig: app_macroscopic_true}). As explained in the main text, this can be linked to the minimum of the diffusion coefficient $\Tilde{\sigma}(V_{\rm{in}})$ (Appendix.~\ref{apsec: Approximate counter dynmaics}) at $V_\text{in} = 0$ seen in \cite{gopal2022large}.  

\end{document}